\documentclass[12pt,letterpaper]{article}

\usepackage{amsmath}
\usepackage{amsfonts}
\usepackage{amssymb}
\usepackage{stmaryrd}
\usepackage{array}
\usepackage[pdftex]{graphicx}
\usepackage{soul}
\usepackage{url}
\usepackage{multicol}
\usepackage{doi}

\newtheorem{ass}{Assumption}

\newtheorem{cor}{Corollary}
\newtheorem{defin}{Definition}

\newtheorem{lem}{Lemma}

\newtheorem{prop}{Proposition}
\newtheorem{rem}{Remark}

\newcommand{\w}{\omega}

\usepackage{tikz}
\usetikzlibrary{positioning,arrows,calc,fit}

\usepackage{calc}
\usepackage{enumitem}
\usepackage{etoolbox}
\newlist{play}{description}{1}
\newlength\widthofactor
\newlength\widthofactors
\newcommand\DeclareActors[1]{
	\renewcommand*\do[1]{
		\csdef{##1}{\item[##1]}
		\setlength\widthofactor{\widthof{##1\quad}}
		\ifdim\widthofactors<\widthofactor
		\setlength\widthofactors\widthofactor
		\fi
	}
	\docsvlist{#1}
	\setlist[play]{labelwidth=\widthofactors, leftmargin=!}
}

\DeclareActors{{All}, E, C, P, N, S}

\begin{document}

\title{\textsc{Implicit Knowledge in Unawareness Structures}\thanks{We thank two anonymous reviewers, the editor, three anonymous reviewers for TARK 2023, Hans van Ditmarsch, participants in TARK 2023 at Oxford University, CSLI Workshop 2022 at Stanford University, and the 2022 Zoom Mini-Workshop on Epistemic Logic with Unawareness for helpful comments. An extended abstract of the paper has been published in the proceedings of TARK 2023. Gaia gratefully acknowledges funding from the Carlsberg Foundation. Burkhard gratefully acknowledges financial support via ARO Contract W911NF2210282.}}
	
\author{Gaia Belardinelli\thanks{Department of Economics, University of California, Davis. Email: gbelardinelli@ucdavis.edu} \and Burkhard C. Schipper\thanks{Department of Economics, University of California, Davis. Email: bcschipper@ucdavis.edu}}
	
\date{May 14, 2024}
	
\maketitle

\maketitle
	
\begin{abstract} Awareness structures by Fagin and Halpern (1988) (FH) feature a syntactic awareness correspondence and accessibility relations modeling implicit knowledge. They are a flexible model of unawareness, and best interpreted from a outside modeler's perspective. Unawareness structures by Heifetz, Meier, and Schipper (2006, 2008) (HMS) model awareness by a lattice of state spaces and explicit knowledge via possibility correspondences. Sublattices thereof can be interpreted as subjective views of agents. Open questions include (1) how implicit knowledge can be defined in HMS structures, and (2) in which way FH structures can be extended to model the agents' subjective views. In this paper, we address (1) by defining implicit knowledge such that it is consistent with explicit knowledge in HMS models. We also introduce a variant of HMS models that instead of explicit knowledge, takes implicit knowledge and awareness as primitives. Further, we address (2) by introducing a category of FH models that are modally equivalent relative to sublanguages and can be interpreted as agents' subjective views depending on their awareness. These constructions allow us to show an equivalence between HMS and FH models. As a corollary, we obtain soundness and completeness of HMS models with respect to the Logic of Propositional Awareness, based on a language featuring \emph{both} implicit and explicit knowledge.
\newline
\newline
\textbf{Keywords:} Unawareness, awareness, implicit knowledge, explicit knowledge.
\newline
\newline
\textbf{JEL-Classifications:} D83, C70.
\bigskip
\end{abstract}
	
\newpage

\begin{flushright}	
Dedicated to Joe Halpern on the occasion of his 70th birthday. 
\end{flushright} 

\section{Introduction}
	
Models of unawareness are of interest to various disciplines, most notably to computer science, economics, game theory, decision theory, 
and philosophy. The seminal contribution in computer science is awareness structures by Fagin and Halpern (1988) (henceforth, \emph{FH models}) who extended Kripke structures with a syntactic awareness correspondence in order to feature notions of implicit knowledge, explicit knowledge, and awareness. In economics, Heifetz, Meier, and Schipper (2006, 2008) introduced unawareness structures (henceforth, \emph{HMS models}) that consist of a lattice of state spaces featuring a notion of explicit knowledge and awareness. Like Kripke structures, HMS models can be constructed canonically and three different sound and complete axiomatizations have been presented (Halpern and R\^{e}go, 2008, Heifetz, Meier, and Schipper, 2008).\footnote{For other approaches and an overview, see Schipper (2015).} There have already been a number of applications to game theory, decision theory, mechanism design and contracting, financial markets, electoral campaigning, conflict resolution, social network formation, business strategy and entrepreneurship etc.\footnote{\emph{Game theory:} See for instance, R\^{e}go and Halpern (2012), Heifetz, Meier, and Schipper (2013b, 2021), Grant and Quiggin (2013), Halpern and R\^{e}go (2014), Meier and Schipper (2014a, 2023), Guarino (2020), Feinberg (2021), Schipper (2021), and Perea (2022). \emph{Decision theory:} See for instance, Karni and Vier{\o} (2013), Schipper (2013, 2014), and Dominiak and Tserenjigmid (2018). \emph{Contract theory and mechanism design:} See for instance, von Thadden and Zhao (2012), Filiz-Ozbay (2012), Auster (2013), Chung and Fortnow (2016), Auster and Pavoni (2024), Francetich and Schipper (2023), and Pram and Schipper (2023). \emph{Financial markets:} See Heifetz, Meier, and Schipper (2013a), Meier and Schipper (2014b), Galanis (2018), Auster and Pavoni (2021), Schipper and Zhou (2021). \emph{Electoral campaigning:} Schipper and Woo (2019). \emph{Conflict resolution:} R\^{e}go and Vieira (2020). \emph{Social network formation:} Schipper (2016). \emph{Business strategy:} Bryan, Ryall, and Schipper (2021).}
	
The different modeling approaches may be seen as reflecting the different foci of the fields. HMS models in economics are very much motivated by game theory and its applications. The main underlying idea is that \emph{explicit} rational reasoning of players is what drives their behavior. Hence, the model features only explicit knowledge (without the detour via implicit knowledge) and awareness, and the sublattices can be interpreted as subjective models of players.\footnote{These subjective views of agents are further elucidated by Grant et al. (2015).} Moreover, the syntax-free frame lends itself seamlessly to the existing body of work in decision theory and game theory.\footnote{For example, Schipper (2014), Heifetz, Meier, and Schipper (2013a), Meier and Schipper (2014a), R\^{e}go and Vieira (2020), Schipper and Zhou (2021), Pram and Schipper (2023).} The focus on behavioral implications also explains why the model is built on strong properties of knowledge such as (positive) introspection and factivity: this allows for the identification of the behavioral implications of unawareness per se without confounding it with mistakes in information processing. Coming from a different angle, FH models were motivated more generally by the study of the logical non-omniscience problem in computer science and philosophy (see e.g., Hintikka, 1975, Levesque, 1984, Lakemeyer, 1986, Stalnaker, 1991). They represent awareness via syntactic awareness correspondences, which for each agent assigns a set of formulas to each state. This approach to awareness modeling offers a great deal of flexibility, because the set of formulas an agent is aware of may be arbitrary, thereby allowing potentially for the representation of different notions of awareness.\footnote{See Fagin and Halpern (1988, pp. 54-55) and Fagin et al. (1995, Chapter 9.5) for discussions.} However, because their semantics is not syntax-free, their applications to decision or game theory require more effort. This is because in decision theory and game theory and applications thereof, the primitives are typically not described syntactically. Moreover, FH models are best interpreted as a tool used by an outside modeler rather than the agents themselves for two reasons: First, the primitive notion of knowledge is \emph{implicit} knowledge while explicit knowledge is derived from implicit knowledge and awareness. Implicit knowledge is not necessarily something that the agent herself can consciously reason about. Second, we cannot think of FH models as models that the agents themselves use for analyzing their epistemic universe unless they are aware of everything. Rather, FH models capture the perspective of a systems designer of a multi-agent distributed system. This becomes relevant in interactive settings when we are interested in the players' interactive perceptions of the epistemic universe. Despite the differences in modeling approaches, Halpern and R\^{e}go (2008) and Belardinelli and Rendsvig (2022) formalize in which ways HMS models are equivalent to FH models in terms of explicit knowledge and awareness. However, as the discussion above makes clear, there remain open questions: First, can implicit knowledge be captured also in HMS models and how would this notion of implicit knowledge be related to implicit knowledge in FH models? Second, can we extend FH models so as to interpret them from the agents' subjective point of views? These questions will be answered in this paper. 

The prior discussion begs a more fundamental question: Why would it even be meaningful to provide answers to the two above questions given the differences in the modeling philosophies between computer science and economics? It is especially interesting to note that there is a notion of knowledge, namely implicit knowledge, that seems to be eschewed in one field but accepted essentially without much discussion in another. Moreover, since the study of awareness is by and large interdisciplinary, we think it is paramount to have a multidisciplinary dialogue exploring the differences between fields addressing the topic and perhaps even resolving these differences. This motivated us to present, as a prelude to our formal study, a literal but fictional dialogue between various disciplines involved in the study of awareness. This allows us to confront with each other the different standpoints taken in these disciplines, illustrate subtle implicit assumptions behind these standpoints, and provide a road map for future work on various notions of implicit knowledge. Envision an economist (E), computer scientist (C), psychologist (P), neuroscientist (N), and sociologist (S) are at dinner together in a restaurant. The economist and computer scientist start a conversation, while others initially chat on a different topic.

\begin{play}
\E I don't understand implicit knowledge in the context of awareness...
	
\C What's the matter? Implicit knowledge begets explicit knowledge when raised into awareness. 

\E Yes, I know that you define explicit knowledge as implicit knowledge and awareness, but what is implicit knowledge?  
	
\C Well, there is nothing mysterious about it. In economics, you are used to model knowledge with partitions in Aumann structures or, as we call it in computer science and logic, indistinguishability relations in Kripke frames. In FH models, implicit knowledge is what is modeled by partitions or more generally accessibility relations. 		

\E Implicit knowledge in FH models looks formally similar to knowledge in Kripke or Aumann structures. However, I think the connotation is quite different. In Aumann or Kripke structures, partitions model explicit knowledge. It is knowledge that the agent herself can consciously reason about rather than just being ascribed to her by the modeler. 
	
\C That's because in Aumann and Kripke structures there is no unawareness. 
	
\E Precisely. Partitions model explicit knowledge by construction. 

\C FH models offer more ``structure'' than Kripke models, which allows us to distinguish between implicit and explicit knowledge, where implicit knowledge is the more primitive concept.

\E Why is implicit knowledge an appropriate primitive of a theory of knowledge under unawareness? Explicit knowledge is what affects her decision-making and thus becomes testable in behavioral data. Why would we want a theory that is based on the untestable construct of implicit knowledge? There is no need to take it as the primitive. As HMS models demonstrate, we can easily take explicit knowledge as the primitive. We can have a testable theory of unawareness. 

\C Hold on, hold on. You seem to be thinking about awareness and knowledge only in relation to \emph{human} decision making, but there are other important contexts as well. For instance, we are very interested in the knowledge of distributed systems.\footnote{Fagin, Halpern, and Vardi (1986) and Halpern (1987).} The system designer may have a pretty good idea of the knowledge implicit in the system and just needs a model to represent it.  

\E Your focus on knowledge systems is interesting. It explains the outside modeler's perspective implicit in FH models.\footnote{This point is made eloquently by Fagin, Halpern, and Vardi (1986): ``The notion of knowledge is \emph{external}. A process cannot answer questions based on its knowledge with respect to this notion of knowledge. Rather, this is a notion meant to be used by the system designer reasoning about the system. ... (I)t does seem to capture the type of intuitive reasoning that goes on by system designers.''} You logicians are quick to endow agents with a formal language but when it comes to models, you keep them stingily for yourself. It is as if all agents that you model live in the pre-Kripke age. That is, you rarely allow your agents to reason with the model herself; models are used by modelers only! Take FH models, could agents use them to analyze their own situation? No, unless everybody is aware of everything.
	
\C Why would agents need models? They reason in a language. Models are for the systems designer to analyze the system. 

\E Well, in game theory we are interested in how the subjective interpretations of a situation drives the agents' interactive behavior.  Standard game theoretic models may be thought of as being shared among players and the modeler. Any player could be a modeler who models various uncertainties and all players' possible state of minds.\footnote{See for instance, Mertens and Zamir (1984).} In games with unawareness using HMS models, we allow each player to be a modeler using her sublattice of spaces of the lattice.\footnote{Meier and Schipper (2014a).} Typically, we only introduce a formal language to investigate the logical foundations of game theoretic models. 

\C Game theoretic modeling is done mainly by analysts. I do not see ordinary people writing game theoretic models when involved in interactions. The lack of subjective interpretation of FH models seems to be just a minor aesthetic issue to me that is of little practical consequence. I also believe we could easily define subjective versions of FH models. Any subjective version of a FH model should be bisimilar to the original one, modulo awareness. Like HMS models, the set of bisimilar FH models would constitute a lattice by set inclusion of atomic formulas. 

\E I would like to see how exactly you construct your subjective versions of FH models. However, the fact that FH models do not have a syntax-free semantics would make them impractical for modeling game theoretic applications.  

\C This seems to be a problem of economists. Working with syntax is very natural in computer science. Again, the lack of syntax-freeness of the semantics of FH models is a minor aesthetic issue that is well overcompensated by the flexibility they provide in modeling various notions of unawareness. And the multiple state spaces of HMS models should make them also more challenging to apply than let's say Kripke frames or Aumann models, shouldn't they?

\E Fair enough. Let's come back to my question: What really is implicit knowledge in FH models? 

\C One way of understanding it is by saying that it is the knowledge that the system has, but has not computed yet. More generally, it is knowledge that is logically implied by what the agent explicitly knows, and that could be reached after some inferential steps. In fact, implicit knowledge was originally introduced in computer science as part of a larger program aimed at addressing the logical omniscience problem.\footnote{Levesque (1984) and Lakemeyer (1986). For logical omniscience problem, see also Hintikka (1975) and Stahlnaker (1991). For more recent work, see Vel\'{a}zquez-Quesada (2013, 2014).} In Kripke structures, knowledge is closed under logical implication. This places unrealistic computational demands on the reasoning abilities of agents. In economics, you talk about bounded rationality since Herbert Simon.\footnote{Simon (1957).} Similarly, in logic we are interested in logical non-omniscience and the distinction between explicit and implicit knowledge allows us to capture some it. 

\E That's not the notion of implicit knowledge in FH models in which awareness is generated by primitive propositions. In those structures, both implicit and explicit knowledge are closed under logical implication, the difference is only the latter is closed only with respect to formulas involving primitive propositions of which the agent is aware. The limits to awareness are fully determined by the primitive propositions the agent is aware of and not by her procedural limits to conceive and reason about events. I do not really see how it is related to Herbert Simon's notions of bounded or procedural rationality, who very much emphasized the limits to the cognitive \emph{processes}.\footnote{Selten (2002).}
	
\C That's correct in so far we limit ourselves to awareness structures featuring awareness generated by primitive propositions. However, awareness structures offer more generally a flexible model in which limits to reasoning processes could be explicitly captured.\footnote{See Fagin and Halpern (1988), pp. 54-55, and Fagin et al. (1995), Chapter 9.5.} 

\E Awareness generated by primitive propositions is the notion of awareness mostly used in applications so far. With such a notion, unawareness is not a form of bounded rationality but rather bounded perception. With respect to what the agent is aware of, she is fully rational when it comes to information processing. So let me rephrase my question: What really is implicit knowledge in FH models with awareness generated by primitive propositions?

\C Implicit knowledge that is not explicit knowledge is knowledge that the system would have if it were aware of it. 
	
\E I don't know what to make of such counterfactual knowledge. If the system were indeed aware of it, then its ``state of mind'' would be obviously different and with it perhaps also its knowledge. What necessitates that the system upon becoming aware would have explicit knowledge that is the same as its pre-aware implicit knowledge? 
	
\C Think of it more as comparative statics exercise that you often do in economics. There are two comparable ``states of mind'' with the same implicit knowledge. They just differ by awareness and hence the explicit knowledge. By studying how the agent's behavior changes when changing \emph{only} awareness, you can isolate the causal effect of awareness keeping knowledge fixed. 
	
\E That would be a very useful exercise indeed. However, FH models do not necessarily model such comparable states of mind. Given a state in an awareness structure, there is not necessarily another state with the same knowledge ``modulo awareness''. This gives me an idea though: In HMS models, we model by construction comparable states of mind that just differ by awareness via projections of states across the lattice of spaces. We can use them to make the comparative statics precise. 
	
\C This would be interesting. Yet, I do not see how you can accomplish this without a notion of implicit knowledge in HMS models. 

\E It should be possible to complement HMS models with a notion of implicit knowledge. However, in contrast to FH models, the notion of implicit knowledge could be a derived notion rather than the primitive notion. 

\C You could define ``implicit'' possibility sets by the inverse image of possibility sets in richer spaces in HMS models. You need to make sure though that such a construction yields a notion of implicit knowledge that preserves properties of ``explicit'' possibility sets such as transitivity and Euclideanness.

\E Indeed. Figuring out the precise conditions under which we can derive implicit knowledge from explicit knowledge is non-trivial. However, it could help us to better understand the notion of implicit knowledge. For instance, a priori it is not clear that there is a unique notion of implicit knowledge that is consistent with explicit knowledge. And this leads me back to the question of whether it is even possible to test different notions of implicit knowledge and empirically identify them in data.

\C I am not sure whether this is necessarily a meaningful question to ask in the context of knowledge of distributed systems. From the construction of the system, the system designer may have a pretty good idea about the way in which implicit knowledge is consistent with the system's explicit knowledge. He designed the system and with it the notion of implicit knowledge. Moreover, the existence of behavioral implications of a notion of knowledge should just be \emph{one} criterion for judging its relevance and properties. If a notion helps us to generate insights, we should study it irrespective of whether or not it has direct behavioral implications. Unobservable primitives have turned out to be useful in economics as well. Just look at utility theory. There you start with unobservable utilities or preferences as primitives. 
	
\E That's why we are interested in revealed preferences. Analogously, what would be the theory of ``revealed implicit knowledge''? How can I conclude from observing a system that it has this or that implicit knowledge? 

\C Since implicit knowledge is knowledge that the system would have if it were more aware, you can raise the system's awareness of some proposition and test it's knowledge about it...
	
\E ...but then I would just test explicit knowledge. Such a test procedure would destroy the very knowledge that I like to test for, namely implicit knowledge that is not explicit yet. That's somewhat reminiscent of Heisenberg's uncertainty principle in quantum physics, where the measurement destroys what is measured. 

\C We do not need to look as far as quantum physics. For instance, even in unawareness structures, it should be nearly impossible to test what the agent is \emph{unaware} of without raising her awareness of the very proposition you want to test for and thereby destroying her unawareness. So already in unawareness structures used in economic applications, unawareness is not testable. 
	
\E That's not the case. We should not assume that whatever an agent chooses must automatically raise her awareness of the contingencies on which the consequences of her choice depend on. In real life, we often make choices whose consequences we cannot fully anticipate. In a choice experiment, by letting an agent choose among contracts that only vary the consequences for a contingency and the negation of that contingency whose awareness I like to test for, I can observe whether the contingency or its negation becomes relevant to the agent's choice. If neither the contingency nor its negation affect choices, then I can conclude from her choices that she must be unaware of the contingency.\footnote{See Schipper (2013) for such a test.} 
	
\C I see, but it also seems to rest on the assumption that implicit knowledge could not affect choices ... 
	
\P Pardon me if I intrude on your conversation, but I overheard you talking about behavioral implications of implicit knowledge. The idea that the unconscious can affect behavior goes back at least to Freud in psychoanalysis (Freud, 1915). In modern psychology, there was a renaissance of the idea in the 1980s in the literature on implicit cognition, priming, and implicit measures.\footnote{See Brownstein et al. (2019) and Gawronski and Hahn (2019).} You might be familiar with it if you took previously an implicit association test.\footnote{Greenwald et al. (1998).} These tests typically reveal some discrimination based on some implicit beliefs of which the individual may not be aware of herself ...
	
\E ... Isn't this literature on implicit measures a part of the replication crisis in psychology? 

\P Indeed, this literature has received a fair share of criticism both in academia and the media. For instance, famous priming studies failed to replicate.\footnote{For instance, Doyen et al. (2012) failed to replicate the famous study by Bargh et al. (1996) in which participants unconsciously primed with age walked slower out of the lab.} Statistical irregularities were discovered.\footnote{See for instance, Schimmack, Heene, and Kesavan (2017)} Reported correlations on the implicit association test decreased over time.\footnote{Schimmack (2021) reevaluates the predictive validity of the Race Implicit Association Test and notes that the reported correlations declined over time from $r = 0.38$ to $r = 0.097$. See also Meissner et al. (2019).} Nevertheless, also statistically small effects can have large effects in society.\footnote{This has been argued with regard to the Implicit Association Test by Greenwald, Banaji, and Nosek (2015).} Moreover, given that there is quite some measurement error in such implicit measures, we shouldn't expect to find large significant effects even if they exist in reality because of the attenuation bias. Finally, there is even disagreement on the very notion of implicit measures, whether they should be understood as measuring automatic biases\footnote{Schimmack (2022)} or unconscious biases\footnote{Gawronski, Ledgerwood, and Eastwick (2022). See also the commentaries on that article in Psychological Inquiry.}. So, I believe more careful research is needed in order to get a better grasp of implicit measures.  
	
\E Hmm... Well, at least the literature offers an idea of how implicit information could have direct behavioral effects without the detour via explicit knowledge. I guess my earlier statement that only explicit knowledge can affect behavior was premature. 
	
\C You mentioned the implicit association test and the implicit biases it might reveal. I was wondering whether an implicit bias doesn't point to mistakes in information processing, as people often disagree completely with their revealed biases. If that's the case, then probably an empirically valid notion of implicit belief should satisfy less stringent properties than explicit belief. 
	
\E That's quite intuitive: If implicit knowledge refers to knowledge that is not necessarily present in the agent's mind and that she could not have revised or discussed with others, then this knowledge may involve mistakes in information processing as captured by violations of reflexivity, transitivity, or Euclideanness of ``implicit'' possibility sets.  In contrast, explicit knowledge is the result of the agent's own reasoning and of discussions she could have had with others. Thus, we should expect explicit knowledge rather than implicit knowledge to satisfy strong properties such as S5.\footnote{``Strong S5 properties'' refer to factivity and positive and negative introspection classically attributed to knowledge; see Rendsvig and Symons (2021).} Explicit knowledge is less prone to mistakes than implicit knowledge simply because the latter lies beneath some level of consciousness so that the individual could not have thought about it carefully.
	
\N 	Well, this runs counter to the received view in neuroscience. The brain is essentially a machine that minimizes consciousness.\footnote{Roth (2003, pp. 236-240)} Processes are automated as much as possible. You do not need to explicitly know everything involved when walking, even in difficult terrain. But you need to consciously reason for instance about a math problem you are trying to solve while walking home! The brain just requires explicit conscious reasoning for ``non-standard'' situations.\footnote{Ibid.} This would suggest that explicit knowledge is prone to mistakes while implicit knowledge is already sophisticated.
	
\E This is another argument for why explicit knowledge and implicit knowledge should not just differ by awareness. The implicit and the explicit ``minds" seem to have different reasoning powers associated to them depending on how we interpret these notions of knowledge. 	

\P Automated implicit knowledge reminds me of Gigerenzer's notion of ecological rationality.\footnote{E.g., Gigerenzer (2002).} Like the ecological rationality of the individual, knowledge may match to the environment frequently encountered by the individual without such knowledge necessarily entering consciousness. But does it necessarily imply that it is sophisticated? If I am not wrong, mistakes in information processing are ruled out by the introspection axioms in the axiomatic system S5. However, these are precisely the axioms that the subject knows that she doesn't know and that she knows that she does know. Doesn't it imply self-consciousness of knowledge? I don't see how it could possibly hold for ecological knowledge or implicit knowledge. Just consider the priming and implicit memory literature\footnote{Roediger and McDermott (1993).}, where subjects are shown quickly some information that they then show to have but not recall having. This clearly violates that subjects know that they know. 
	
\C Not necessarily. This just shows that they do not \emph{explicitly} know what they know implicitly. But they could still implicitly know that they know it implicitly. 

\P Hmm, this reminds me of studies that showed that people were able to predict their own implicit bias pretty well.\footnote{Hahn et al. (2014).} This would suggest that in some sense they are aware of their implicit bias and that its not an unconscious phenomenon. 
	
\S To me it sounds like implicit knowledge could be interpreted as tacit knowledge \`{a} la Michael Polanyi\footnote{Polanyi (1962).} Tacit knowledge is knowledge that the individual has, but she doesn't have the words to express. In this sense it is implicit. 
	
\C That's an interesting interpretation of implicit knowledge.
	
\E How could you test for what the individual knows only tacitly? 
	
\S Tacit knowledge affects behavior. An opera star obviously knows how to sing but she may not necessarily be able to describe how to sing. Similarly, Stradivari knew how to build excellent violins but was apparently unable to explicitly teach it to his successors.
	
\N I cannot help noticing that in these examples the implicit tacit knowledge is the more sophisticated notion rather than the explicable knowledge. That conforms very much to how we think about implicit routine knowledge in neuroscience. 
	
\E  To me the interpretation of tacit knowledge differs from implicit knowledge. Explicit knowledge is distinguished from implicit knowledge by awareness (at least in the formal literature on awareness), while ``explicated'' knowledge is apparently distinguished from tacit knowledge by the ability to explicate or communicate it, not by awareness.
	
\C I wonder whether at a formal level, a theory of ``explicability'' would look very different from a theory of awareness. After all, awareness generated by primitive propositions is defined by the set of primitive propositions on which the agent's language is built. So implicit knowledge may also be interpretable as knowledge that the agent has but doesn't have the language to express, i.e., tacit knowledge.
	
\E Well, all those notions of implicit knowledge seem to have in common that they are beneath (or better beyond) some level of awareness or explicability. Perhaps HMS models are a useful tool to study them because the lattice structure reflects levels of awareness/explicability. By fitting them into decision and game theoretic models, we could develop rigorous testable theories of tacit knowledge and of implicit biases for instance. This would be a relevant enterprise. 
	
\C To implicit knowledge!
	
\All Cheers!
	
\end{play}

The dialogue illustrates the differences between FH and HMS models and, more generally, the less obvious differences in the ``modeling cultures'' of computer science and economics. At the same time, it suggests that the types of knowledge that arise in these structures are interesting and relevant to various disciplines beyond economics and computer science. Finally, it outlines a multidisciplinary agenda of studying versions of implicit knowledge that go beyond the extant notion of implicit knowledge in FH models. As a first step, the dialogue motivates a reconciliation of the two approaches, a step we undertake in this paper. Both approaches can be extended to encompass features of the other. As suggested informally in the dialogue, we extend HMS models to implicit knowledge and FH models to categories encompassing their subjective interpretations by agents. These extensions enable us to show an equivalence of the two approaches also with respect to implicit knowledge and provide sound and complete axiomatizations of HMS models also with respect to implicit knowledge. However, the dialogue embeds this first step into a broader research program on implicit knowledge. It motivates a comprehensive study of notions of implicit knowledge behind some veil of expressibility that are potentially of use to other disciplines. While we cannot deliver such a comprehensive study in a single paper, in follow-up work, quasi as a second step, we introduce a notion of tacit knowledge in HMS models and discuss various applications (Belardinelli and Schipper, 2024). 

We define implicit knowledge in HMS models so that it is consistent explicit knowledge, thereby providing a way to understand implicit knowledge in terms of explicit knowledge in this paper. We are aware of only a few other approaches deriving implicit knowledge from explicit knowledge. Using neighborhood models without a notion of awareness, Vel{\'a}zquez-Quesada (2013) takes explicit knowledge as the primitive and then derives implicit knowledge as closure of logical consequences of explicit knowledge. Implicit knowledge is then understood as knowledge that the agent ideally could deduce from her explicit knowledge. Lorini (2020) takes an agent's belief base as explicit knowledge and derives implicit knowledge as what is deducible from an agent's belief base and common background information. While we find these two notions of implicit knowledge easy to interpret, it is not the notion of implicit knowledge that is captured by FH models when awareness is generated by a set of primitive propositions, as argued by the economist in the dialogue. We also introduce a variant of HMS models in which we take the notion of implicit knowledge and a semantic awareness function as the primitive, and then derive explicit knowledge (we call these \emph{implicit-knowledge based HMS models}). This shows that in HMS models, implicit and explicit knowledge are ``interdefinable'', at least in the sense that taking any of the two as primitive allows us to consistently specify the other, so one may choose either one as a primitive. 

We are also interested in an extension of FH models that allows us to interpret them as subjective views of agents. Starting from an FH model, we show how to obtain such subjective views by forming a category of FH models with FH models as objects and surjective bounded morphisms as morphisms. Each category of FH models is literally a category of FH models that are modally equivalent relative to sublanguages formed by taking subsets of atomic formulas. The category of FH models forms a complete lattice ordered by subset inclusion on sets of atomic formulas or ordered by surjective bounded morphisms or ordered by modal equivalence relative to sublanguages. Each FH model in the lattice can be interpreted as the subjective model of an agent where her awareness level is given by the subset of atomic formulas over which the FH model is defined. The construction now suggests transformations between FH and HMS models. The transformation from FH to HMS models relies on a transformation of each FH category into an implicit knowledge-based HMS model mentioned above. This implicit knowledge-based HMS model can be complemented with explicit knowledge and thus yields a HMS model. The transformation from HMS to FH models simply relies on pruning away the subjective spaces, only maintaining the upmost space in the lattice, as well as deriving the syntactic awareness correspondences from possibility correspondences and the lattice of spaces in HMS models. For each model class, its transformation into a model of the other class satisfies the same formulas from a language for explicit, implicit knowledge, and awareness. As a corollary of soundness and completeness of the Logic of Propositional Awareness with respect to the class of FH models, the results allow us to derive soundness and completeness for the class of HMS models \emph{now with implicit knowledge}, complementing earlier axiomatizations of HMS models that made use of explicit knowledge (and awareness) only (Heifetz, Meier, and Schipper, 2008, Halpern and R\^{e}go, 2008).

\section{HMS Models}

HMS models are multi-agent models for awareness originally proposed by Heifetz, Meier, and Schipper (2006, 2008). The goal was to provide a syntax-free semantics of multi-person unawareness that would fit seamlessly in decision and game theory. The model also served as an answer to the so-called impossibility result by Dekel, Lipman, and Rustichini (1998), who showed that state space models cannot capture non-trivial unawareness. HMS models recalled in this section escape the triviality by considering a lattice of spaces while FH models discussed in later sections do so by adding a syntactic awareness correspondence for each agent.

Throughout the paper, we let $\mathsf{At}$ be a non-empty set of atomic formulas.
	
\begin{defin}[HMS Model]\label{HMS_model}  A \emph{HMS model} $\mathsf{M} = \langle I, \{S_{\Phi}\}_{\Phi \subseteq \mathsf{At}}, (r^{\Phi}_{\Psi})_{\Psi \subseteq \Phi \subseteq \mathsf{At}}, (\Pi_i)_{i \in I}, v \rangle$ for $\mathsf{At}$ consists of

\begin{itemize}
			
	\item a non-empty set of individuals $I$,
			
	\item a non-empty collection of non-empty disjoint state spaces $\{S_{\Phi}\}_{\Phi \subseteq \mathsf{At}}$ indexed by subsets of atomic formulas $\Phi \subseteq \mathsf{At}$. Note that $\{S_{\Phi}\}_{\Phi \subseteq \mathsf{At}}$ forms a complete lattice by subset inclusion on the set of atomic formulas $\Phi \subseteq \mathsf{At}$. We write $S_{\Phi'} \succeq S_{\Phi}$ if $\Phi \subseteq \Phi' \subseteq \mathsf{At}$ and say that $S_{\Phi'}$ is more expressive than $S_{\Phi}$. Denote the set of all states in spaces of the lattice by $\Omega := \bigcup_{\Phi \subseteq \mathsf{\mathsf{At}}} S_{\Phi}$.
			
	\item Projections $(r^\Phi_\Psi)_{\Psi \subseteq \Phi \subseteq \mathsf{At}}$ such that, for any $\Phi, \Psi \subseteq \mathsf{\mathsf{At}}$ with $\Psi \subseteq \Phi$, $r^\Phi_\Psi: S_{\Phi} \longrightarrow S_{\Psi}$ is surjective, for any $\Phi \subseteq \mathsf{\mathsf{At}}$, $r^{\Phi}_{\Phi}$ is the identity on $S_{\Phi}$, and for any $\Phi, \Psi, \Upsilon \subseteq \mathsf{\mathsf{At}}$, $\Upsilon \subseteq \Psi \subseteq \Phi$, $r^{\Phi}_{\Upsilon} = r^{\Psi}_{\Upsilon} \circ r^{\Phi}_{\Psi}$. For any $\Phi \subseteq \mathsf{At}$ and $D \subseteq S_{\Phi}$, denote by $D^{\uparrow} := \bigcup_{\Phi \subseteq \Psi \subseteq \mathsf{At}} (r^{\Psi}_{\Phi})^{-1}(D)$. An \emph{event} $E \subseteq \Omega$ is defined by a subset $\Phi \subseteq \mathsf{At}$ and a subset $D \subseteq S_{\Phi}$ such that $E := D^{\uparrow}$. We call $S_{\Phi}$ the base-space of the event $E$ and $D$ the base of the event $E$. We denote by $\Sigma$ the set of events.
			
	\item A possibility correspondence $\Pi_i: \Omega \longrightarrow 2^{\Omega} \setminus \{\emptyset\}$ for each individual $i \in I$.
			
	\item A valuation function $v: \mathsf{At} \longrightarrow \Sigma$.
			
\end{itemize}
\end{defin}
	
While Heifetz, Meier, and Schipper (2006) allowed for any non-empty complete lattice of non-empty disjoint spaces, we consider a lattice in which spaces are indexed by subsets of atomic formulas $\Phi \subseteq \mathsf{At}$ and the lattice order is induced by subset inclusion on the set of atomic formulas. This anticipates the connection to syntax (introduced by Heifetz, Meier, and Schipper, 2008) used in later sections. Implicitly, we can consider states in a space $S_{\Phi}$ as consistent subsets of formulas in a language formed from the subset of atomic formulas in $\Phi \subseteq \mathsf{At}$. More expressive spaces in the lattice are associated with richer sets of atomic formulas and so richer sublanguages. The partial order between the state spaces induced by subset inclusion is thus also an order between expressiveness of languages. We refer for further discussions to Heifetz, Meier, and Schipper (2006, 2008).

Not every subset of the union of spaces is an event. Intuitively, $D^\uparrow$ collects all the ``extensions of descriptions in $D$ to at least as expressive vocabularies" (Heifetz, Meier, and Schipper, 2006). Events are well defined by the above definition except for the case of vacuous events. Since the empty set is a subset of any space, we have as many vacuous events as there are state spaces. These vacuous events are distinguished by their base-space, so we denote them by $\emptyset^{S_{\Phi}}$ for $\Phi \subseteq \mathsf{At}$. At a first glance, the existence of many vacuous events may seem puzzling. Note that vacuous events essentially represent contradictions, i.e., propositions that cannot hold at any state. Contradictions are formed with atomic formulas. Thus, they can be more or less complicated depending on the expressiveness of the underlying language describing states and hence are represented by different vacuous events.
	
We define Boolean operations on events. Negation of events is defined as follows: Let $E$ be an event with base $D$ and base-space $S_{\Phi}$. Then $\neg E := (S_{\Phi} \setminus D)^{\uparrow}$. Conjunction of events is defined by intersection of events. Disjunction of events is defined by the DeMorgan Law using negation and conjunction as just defined. Note that in HMS models we typically have $E \cup \neg E \subsetneqq \Omega$ unless the base-space of $E$ is $S_{\emptyset}$, the meet of the lattice of spaces. Also, disjunction of two events is typically a proper subset of the union of these events unless both events have the same base-space, since it is just the union of the events in spaces in which both events are ``expressible''.
	
The following notation will be convenient: Sometimes we denote by $S_{\omega}$ the state space that contains state $\omega$.  For any $D \subseteq S_{\Phi}$, we denote by $D_{S_{\Psi}}$ the projection of $D$ to $S_{\Psi}$ for $\Psi \subseteq \Phi \subseteq \mathsf{At}$. We simplify notation further and let for any $D \subseteq S_{\Phi}$ and $\Psi \subseteq \Phi \subseteq \mathsf{At}$, $D_{\Psi}$ be the projection of $D$ to $S_{\Psi}$. Similarly, for any $D \subseteq S_{\Psi}$, we denote by $D^{\Phi}$ the ``elaboration'' of $D$ in the space $S_{\Phi}$ with $\Psi \subseteq \Phi$, i.e., $D^{\Phi} := (r^{\Phi}_{\Psi})^{-1}(D)$. The same applies to states, i.e., $\omega_{\Psi}$ is the projection of $\omega \in S_{\Phi}$ to $S_{\Psi}$ with $\Psi \subseteq \Phi$. Finally, for any event $E \in \Sigma$, we denote by $S(E)$ the base-space of $E$. We say that an event $E$ is expressible in $S_\Phi$ if $S(E) \preceq S_\Phi$. 
	
As usual in epistemic structures used in game theory and economics, information is modeled by a possibility correspondence instead of an accessibility relation. In HMS models, having mappings rather than relations adds extra convenience in that we can easily compose projections with possibility correspondences and vice versa. It is precisely the projective structure that makes HMS models tractable in applications and lets us analyze phenomena across ``awareness levels'' $\{S_{\Phi}\}_{\Phi \subseteq \mathsf{At}}$. Since the motivation for HMS models in game theory and economics is to isolate the behavioral implications of unawareness from other factors like mistakes in information processing etc., we require that possibility correspondences satisfy strong properties analogous to S5.\footnote{Again, we refer to Heifetz, Meier, and Schipper (2006, 2008) for discussions of these properties. Generalizations are considered by Heifetz, Meier, and Schipper (2013a), Halpern and R\^{e}go (2008), Board, Chung, and Schipper (2011), and Galanis (2011, 2013).} 
	
\begin{ass}[Properties of the Possibility Correspondence]\label{assumptions_HMS} For any individual $i \in I$, we require that the possibility correspondence $\Pi_i$ satisfies
\begin{itemize}

\item[] \emph{Confinement:} If $\omega \in S_{\Phi}$, then $\Pi_{i}(\omega)\subseteq
			S_{\Psi}$ for some $\Psi \subseteq \Phi$.
			
\item[] \emph{Generalized Reflexivity:} $\omega \in \Pi_{i}^{\uparrow}(
			\omega)$ for every $\omega \in \Omega$.\footnote{Here and in what follows, we abuse notation slightly and write $\Pi_{i}^{\uparrow }(\omega)$ for $\left(\Pi _{i}(\omega )\right)^{\uparrow }$.}
			
\item[] \emph{Stationarity:} $\omega' \in \Pi_{i}(\omega)$
			implies $\Pi_{i}(\omega') =\Pi_{i}(\omega)$.
			
\item[] \emph{Projections Preserve Ignorance:} If $\omega \in S_{\Phi}$ and $\Psi \subseteq \Phi$, then $\Pi_{i}^{\uparrow}(\omega) \subseteq \Pi_{i}^{\uparrow}(\omega_{\Psi})$.
			
\item[] \emph{Projections Preserve Knowledge:} If $\Upsilon \subseteq \Psi \subseteq \Phi$, $\omega \in S_{\Phi}$ and $\Pi_{i}(\omega)\subseteq S_{\Psi}$ then $\left(\Pi_{i}( \omega)\right)_{\Upsilon} = \Pi_{i}(\omega_{\Upsilon})$.
\end{itemize}
\end{ass}

Sometimes we denote by $S_{\Pi_i(\omega)}$ the state space $S$ for which $\Pi_i(\omega) \subseteq S$. 

Stationarity corresponds to both transitivity and Euclideaness, while together Generalized Reflexivity and Stationarity corresponds to partitional properties in Kripke frames or Aumann structures. Note though that $\Pi_i$ does not necessarily form a partition of a state space because we allow $S_{\Pi_i(\omega)} \preceq S_{\omega}$. We refer to Heifetz, Meier, and Schipper (2006, 2008) for further discussions of these properties.

Given the possibility correspondence, the knowledge operator is essentially defined like in Aumann (1999). 

\begin{defin}[Knowledge Operator]\label{definK} For every individual $i \in I$, the \emph{knowledge operator} on events is defined by, for every event $E \in \Sigma$,
	\begin{eqnarray*} K_i(E) & := & \{\omega \in \Omega : \Pi_i(\omega) \subseteq E \}
	\end{eqnarray*} if there exists a state $\omega \in \Omega$ such that $\Pi_i(\omega) \subseteq E$, and by $K_i(E) = \emptyset^{S(E)}$ otherwise.
\end{defin} 
\begin{defin}[Awareness Operator]\label{definA} For every individual $i \in I$, the \emph{awareness operator} on events is defined by, for every event $E \in \Sigma$,
	\begin{eqnarray*} A_i(E) & := & \{\omega \in \Omega : S_{\Pi_i(\omega)} \succeq S(E) \}
	\end{eqnarray*} if there exists a state $\omega \in \Omega$ such that $S_{\Pi_i(\omega)} \succeq S(E)$, and by $A_i(E) = \emptyset^{S(E)}$ otherwise. The unawareness operator is defined by $U_i(E) := \neg A_i(E)$.
\end{defin}

We read $K_i(E)$ as ``individual $i$ knows the event $E$'' and $A_i(E)$ as ``individual $i$ is aware of event $E$''. 

\begin{lem}[Heifetz, Meier, and Schipper, 2006]\label{alles_events} For every individual $i \in I$ and event $E \in \Sigma$, both $K_i(E)$ and $A_i(E)$ are $S(E)$-based events.
\end{lem}
	
\begin{prop}[Heifetz, Meier, and Schipper, 2006]\label{strong_explicit_knowledge} For every individual $i \in I$, the knowledge operator $K_i$ satisfies the following properties: For every $E, F \in \Sigma$ and $\{E_n\}_n \subseteq \Sigma$,
\begin{itemize}
\item[(i)] Necessitation: $K_{i}(\Omega) = \Omega$,
			
\item[(ii)] Conjunction: $K_{i}\left( \bigcap_{n} E_{n}\right) = \bigcap_{n}K_{i}\left(E_{n}\right)$,
			
\item[(iii)] Truth: $K_{i}(E)\subseteq E$,
			
\item[(iv)] Positive Introspection: $K_{i}(E)\subseteq K_{i}K_{i}(E)$,
		
\item[(v)] Monotonicity: $E \subseteq F$ implies $K_{i}(E)\subseteq K_{i}(F)$.
			
\item[(vi)] Weak Negative Introspection I: $\neg K_{i}(E)\cap \neg K_{i} \neg K_{i}(E) \subseteq \neg K_{i} \neg K_{i} \neg K_{i}(E)$.
\end{itemize}
\end{prop}
	
\begin{prop}[Heifetz, Meier, and Schipper, 2006]\label{prop: K-U properties} For every individual $i \in I$, the following properties of knowledge and awareness obtain: For every $E \in \Sigma$ and $\{E_n\}_n \subseteq \Sigma$,
\begin{enumerate}
\item $KU$ Introspection: $K_{i}U_{i}(E) = \emptyset ^{S(E)}$,
			
\item $AU$ Introspection: $U_{i}(E) = U_{i}U_{i}(E)$
			
\item Weak Necessitation: $A_{i}(E) = K_{i}(S(E)^{\uparrow})$,
			
\item Plausibility: $A_{i}(E) = K_{i}(E) \cup K_{i} \lnot K_{i}(E)$,
			
\item Strong Plausibility: $U_{i}(E) = \bigcap_{n=1}^{\infty}\left(\lnot K_{i}\right)^{n}(E)$,
			
\item Weak Negative Introspection II: $\lnot K_{i}(E)\cap A_{i}\lnot K_{i}(E) = K_{i}\lnot K_{i}(E)$,
			
\item Symmetry: $A_{i}(E) = A_{i}(\lnot E)$,
			
\item $A$-Conjunction: $\bigcap_{n} A_{i}\left(E_{n}\right) = A_{i}\left(\bigcap_{n} E_{n}\right)$,
			
\item $AK$-Self Reflection: $A_{i}(E) = A_{i}K_{i}(E)$,
			
\item $AA$-Self Reflection: $A_{i}(E) = A_{i}A_{i}(E)$,
			
\item $A$-Introspection: $A_{i}(E) = K_{i}A_{i}(E)$.
\end{enumerate}
\end{prop}

Heifetz, Meier, and Schipper (2006, 2008) also define mutual knowledge, common knowledge, mutual awareness, and common awareness operators and derive properties, which we skip here for brevity.

The following lemma turns out to be very useful but has not been proved in the literature. It proves a consistency condition of the possibility correspondence. To see the content of the lemma, consider any state $\omega$ and its projection (``less expressive version'') $\omega_{\Psi}$. If at $\omega$ the individual's knowledge and awareness is given by $\Pi_i(\omega) \subseteq S_{\Upsilon}$, for some $S_{\Upsilon}$ that involves even less awareness than $S_{\Psi}$ (i.e., where $\Upsilon\subseteq \Psi$), then at $\omega_{\Psi}$ the individual's knowledge and awareness is also given by $\Pi_i(\omega)$. That is, at a less expressive projection $\omega_{\Psi}$, the individual cannot be aware of more than she is at $\omega$ and cannot have knowledge that differs from her knowledge at $\omega$.

\begin{lem}\label{lemma: pc in comparable spaces are equal} For every individual $i \in I$ and any $\Upsilon \subseteq \Psi \subseteq \Phi \subseteq \mathsf{At}$, if $\omega \in S_{\Phi}$ and $\Pi_i(\omega) \subseteq S_\Upsilon$, then $\Pi_i(\omega_\Psi) = \Pi_i(\omega)$.
\end{lem}

\noindent \textsc{Proof. } By Projections Preserve Ignorance, $\Pi_i^{\uparrow}(\omega_{\Psi}) \subseteq \Pi_i^{\uparrow}(\omega_{\Upsilon})$ since $(\omega_{\Psi})_{\Upsilon} = \omega_{\Upsilon}$. By Generalized Reflexivity, $\omega_\Upsilon \in \Pi_i(\omega)$. By Stationarity, $\Pi_i(\omega_{\Upsilon}) = \Pi_i(\omega)$. Thus, $\Pi_i^{\uparrow}(\omega_{\Psi}) \subseteq \Pi_i^{\uparrow}(\omega)$.

Also, if $\Pi_i(\omega_{\Psi}) \subseteq S_{\Upsilon}$, then $\Pi_i(\omega_{\Psi}) = \Pi_i(\omega)$ follows from Stationarity. It now follows that if $\Pi_i(\omega_{\Psi}) \neq \Pi_i(\omega)$, we must have $\Pi_i(\omega_{\Psi}) \subseteq S_{\Delta}$ with $\Upsilon \subsetneqq \Delta \subseteq \Psi$. Then there exists an event $E \in \Sigma$ with $S(E) = S_\Upsilon$ s.t. $\omega_{\Psi} \in A_i(E)$ and $\omega \in U_i(E)$. Since $U_i(E)$ is an $S(E)$ based event by Lemma~\ref{alles_events} and the definition of negation, we must have $\omega_{\Psi} \in U_i(E)$, a contradiction to $\omega_{\Psi} \in A_i(E)$. \hfill $\Box$

\section{From Explicit to Implicit Knowledge\label{explicit-to-implicit}}
	
In this section, we define the ``implicit'' possibility correspondence $\Lambda_i$ such that it is consistent with the ``explicit'' possibility correspondence $\Pi_i$. We then define implicit knowledge as based on $\Lambda_i$ and show that it satisfies standard S5 properties as well as properties of Fagin and Halpern (1988) that are jointly satisfied by implicit knowledge, explicit knowledge, and awareness. 

From now on, for any individual $i \in I$, we call $\Pi_i$ the \emph{explicit} possibility correspondence, $\Pi_i(\w)$ \emph{explicit} possibility set at $\omega$, and $K_i(E)$ the event that $i$ \emph{explicitly} knows $E$. 
	
\begin{defin}[Implicit Possibility Correspondence] Given the explicit possibility correspondence $\Pi_i$ of individual $i \in I$, let the \emph{implicit possibility correspondence} $\Lambda_i: \Omega \longrightarrow 2^\Omega$ satisfy
\begin{itemize}
\item[] \emph{Reflexivity:} For any $\omega \in \Omega$, $\omega \in \Lambda_i(\omega)$.

\item[] \emph{Stationarity}: $\omega' \in \Lambda_i(\omega)$ implies $\Lambda_i(\omega') = \Lambda_i(\omega)$.

\item[] \emph{Projections Preserve Implicit Knowledge:} For any $\Phi \subseteq \mathsf{At}$, if $\omega \in S_{\Phi}$, then $\Lambda_i(\omega)_{\Psi} = \Lambda_i(\omega_{\Psi})$ for all $\Psi \subseteq \Phi$.

\item[] \emph{Explicit Measurability:} $\omega' \in \Lambda_i(\omega)$ implies $\Pi_i(\omega') = \Pi_i(\omega)$.

\item[] \emph{Implicit Measurability:} $\omega' \in \Pi_i(\omega)$ implies $\Lambda_i(\omega') = \Lambda_i(\omega)_{S_{\Pi_i(\omega)}}$.
\end{itemize} 
\end{defin}

\begin{defin}[Complemented HMS Model] Given an HMS model $M = \langle I, \{S_{\Phi}\}_{\Phi \subseteq \mathsf{At}},\\(r^{\Phi}_{\Psi})_{\Psi \subseteq \Phi \subseteq \mathsf{At}}, (\Pi_i)_{i \in I}, v \rangle$ and a collection of implicit possibility correspondences $(\Lambda_i)_{i \in I}$ satisfying the above properties, we call $\overline{\mathsf{M}} = \langle I, \{S_{\Phi}\}_{\Phi \subseteq \mathsf{At}},(r^{\Phi}_{\Psi})_{\Psi \subseteq \Phi \subseteq \mathsf{At}}, (\Pi_i)_{i \in I}, (\Lambda_i)_{i \in I}, v \rangle$ a \emph{complemented HMS model}.
\end{defin}

A complemented HMS model is a HMS model complemented with an implicit possibility correspondence for each individual. In the following, we discuss and derive properties of the implicit possibility correspondence. It also demonstrates ways in which the implicit possibility correspondence is consistent with the explicit possibility correspondence. 

Reflexivity and Stationarity are standard and imply that $\{\Lambda_i(\omega)\}_{\omega \in S_{\Phi}}$ forms a partition of $S_{\Phi}$ for every $\Phi \subseteq \mathsf{At}$. It is straightforward to see that they also imply a strengthening of Confinement: The implicit possibility set at a state must be a subset of the space in which the state itself lies. That is, both the state and the implicit possibility set are described using the \emph{same} language. More formally:

\begin{rem}[Strong Confinement] For any individual $i \in I$, $\Phi \subseteq \mathsf{At}$, and $\omega \in S_{\Phi}$, $\Lambda_i(\omega) \subseteq S_{\Phi}$.
\end{rem}

Projections Preserve Implicit Knowledge is analogous to Projections Preserve Knowledge satisfied by $\Pi_i$. The absence of Projections Preserve (Implicit) Ignorance from the above list of properties imposed to the implicit possibility correspondence may look puzzling at a first glance. Yet, as we show below it is implied by Strong Confinement and Projections Preserve Implicit Knowledge.
	
\begin{lem}[Projections Preserve Implicit Ignorance] For any individual $i \in I$, if $\Lambda_i$ satisfies Strong Confinement and Projections Preserve Implicit Knowledge, then $\Lambda_i$ satisfies Projections Preserve Implicit Ignorance. That is, for all $\Phi \subseteq \mathsf{At}$, if $\omega \in S_{\Phi}$, then $\Lambda_i^{\uparrow}(\omega) \subseteq \Lambda_i^{\uparrow}(\omega_{\Psi})$ for all $\Psi \subseteq \Phi$.
\end{lem}

\noindent \textsc{Proof. } If $\omega \in S_{\Phi}$, then $\Lambda_i(\omega) \subseteq S_{\Phi}$ by Strong Confinement. By Projections Preserve Implicit Knowledge, $(\Lambda_i(\omega)_{\Psi})^{\uparrow} = \Lambda_i^{\uparrow}(\omega_{\Psi})$. Since $\Psi \subseteq \Phi$, it follows now that $\Lambda_i^{\uparrow}(\omega) \subseteq \Lambda_i^{\uparrow}(\omega_{\Psi})$. \hfill $\Box$\\

\begin{figure}\caption{Two Single-Agent Examples of Complemented HMS Models\label{examples}}
\begin{center}
\includegraphics[scale = 0.08]{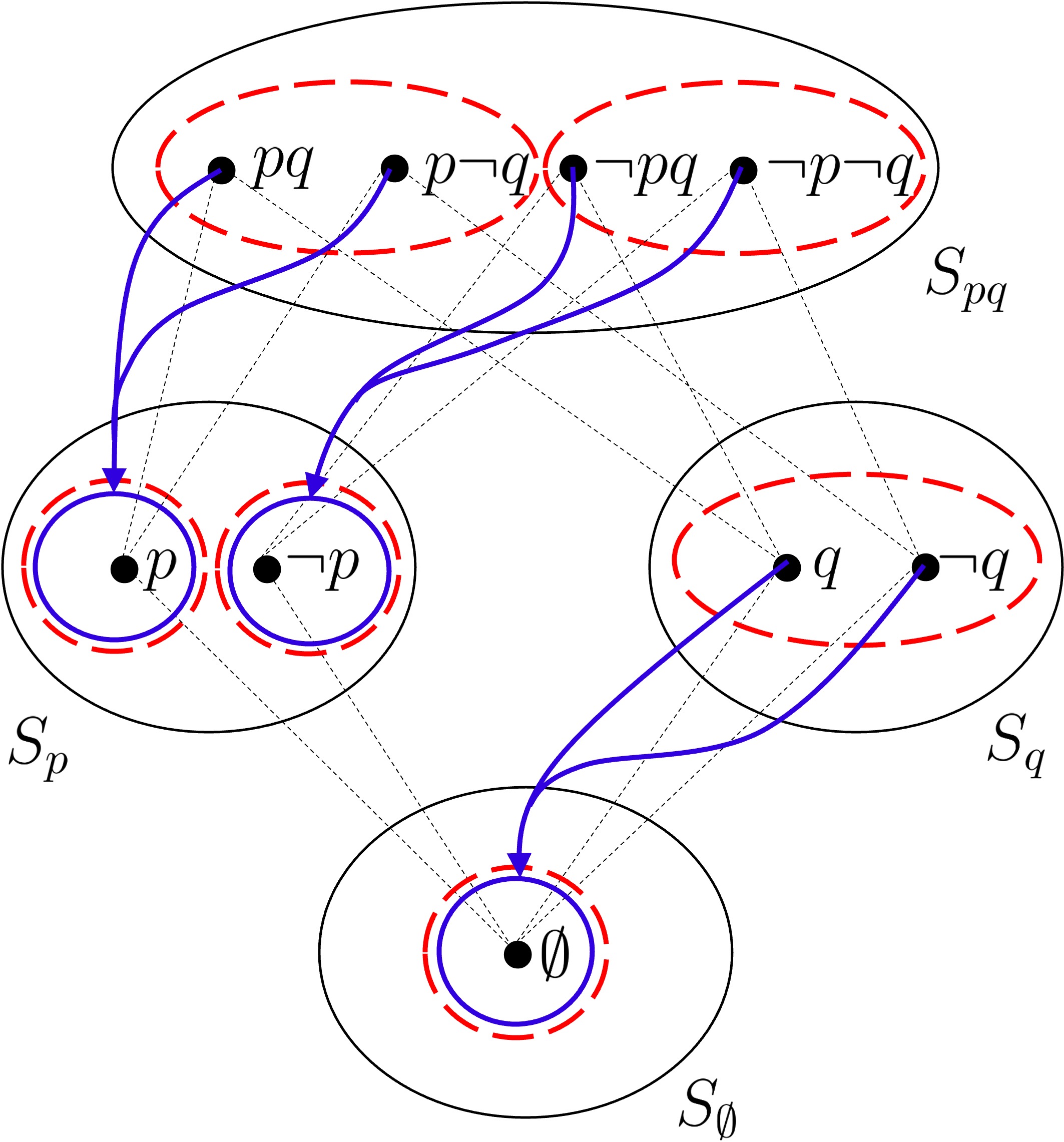} \quad \quad \quad \includegraphics[scale = 0.08]{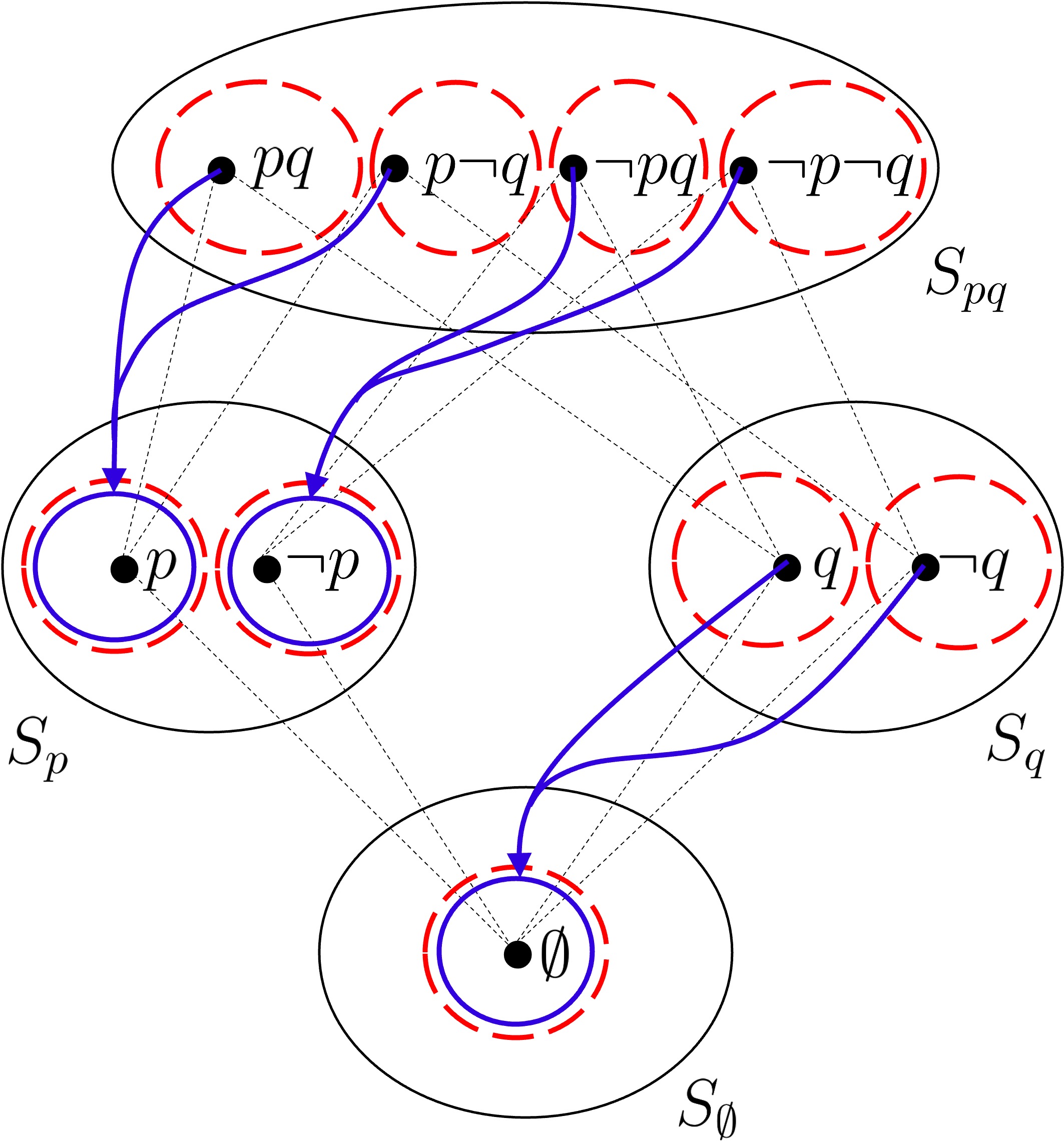}
\end{center}
\end{figure}

Explicit Measurability says that explicit knowledge is measurable with respect to implicit knowledge. That is, the agent always implicitly knows her explicit knowledge. The converse, Implicit Measurability, is more subtle because of awareness. An individual may not explicitly know her implicit knowledge because she might be unaware of some events. However, the individual always explicitly knows her implicit knowledge modulo awareness. That is, she might implicitly know more at a higher awareness level than what she knows at her awareness level (like in the structure to the right in Figure~\ref{examples}) but at her awareness level, her implicit knowledge equals her explicit knowledge. The following lemma formalizes the last conclusion. The proof uses all properties of $\Pi_i$ and $\Lambda_i$ except Projections Preserve Knowledge of both $\Lambda_i$ and $\Pi_i$ and Projections Preserve Ignorance of $\Pi_i$.

\begin{lem}\label{coincides} For any individual $i \in I$, if $\omega' \in \Pi_i(\omega)$, then $\Lambda_i(\omega') = \Pi_i(\omega')$.
\end{lem}

\noindent \textsc{Proof.} Let $\omega' \in \Pi_i(\omega)$. This implies $\Pi_i(\omega') = \Pi_i(\omega)$ by Stationarity of $\Pi_i$ and $\Lambda_i(\omega') = \Lambda_i(\omega)_{S_{\Pi_i(\omega)}}$ by Implicit Measurability. By Generalized Reflexivity of $\Pi_i$, we have $\omega' \in \Pi_i(\omega')$ and by Reflexivity of $\Lambda_i$, we have $\omega' \in \Lambda_i(\omega')$. Thus, $\Pi_i(\omega') \cap \Lambda_i(\omega') \neq \emptyset$.

Suppose that there exists $\omega'' \in \Pi_i(\omega')$ with $\omega'' \notin \Lambda_i(\omega')$. Then by Stationarity of $\Lambda_i$, $\Lambda_i(\omega'') \cap \Lambda_i(\omega') = \emptyset$. But $\omega'' \in \Pi_i(\omega')$ implies in this case by Implicit Measurability that $\Lambda_i(\omega'') = \Lambda_i(\omega)$, a contradiction.

Now suppose that there exists $\omega'' \in \Lambda_i(\omega')$ with $\omega'' \notin \Pi_i(\omega')$. By Stationarity of $\Pi_i$, $\Pi_i(\omega'') \cap \Pi_i(\omega') = \emptyset$. But $\omega'' \in \Lambda_i(\omega')$ implies by Explicit Measurability that $\Pi_i(\omega'') = \Pi_i(\omega')$, a contradiction. \hfill $\Box$\\

Above properties imply now a strong connection between implicit and explicit possibility sets:

\begin{lem}[Coherence]\label{coherence} For any individual $i \in I$, $\omega \in \Omega$, $\Lambda_i(\omega)_{S_{\Pi_i(\omega)}} = \Pi_i(\omega)$.
\end{lem}

\noindent \textsc{Proof.} For all $\omega \in \Omega$, by Confinement and Generalized Reflexivity, $\omega_{S_{\Pi_i(\omega)}} \in \Pi_i(\omega)$. By Lemma~\ref{coincides}, $\Lambda_i(\omega_{S_{\Pi_i(\omega)}}) = \Pi_i(\omega_{S_{\Pi_i(\omega)}})$. By Stationarity of $\Pi_i$, $\Pi_i(\omega_{S_{\Pi_i(\omega)}}) = \Pi_i(\omega)$. Thus, $\Lambda_i(\omega_{S_{\Pi_i(\omega)}}) = \Pi_i(\omega)$. By Projections Preserve Implicit Knowledge of $\Lambda_i$, $\Lambda_i(\omega)_{S_{\Pi_i(\omega)}} = \Pi_i(\omega)$.\hfill $\Box$\\
		
Figure~\ref{examples} shows two single-agent examples of how implicit knowledge is fitted to explicit knowledge. Consider first the HMS model to the left. There are four spaces indexed by subsets of atomic formulas. Anticipating the semantics of HMS models introduced later, we describe and call states by their atomic formulas. The explicit possibility correspondence of the individual is indicated by the solid blue ovals and arrows. For instance, at state $pq$ she considers possible state $p$. That is, she is unaware of $q$ and knows $p$. Similarly, at state $\neg p q$ she is unaware of $q$ and knows $\neg p$. Her implicit possibility correspondence is given by the red dashed ovals. Note that in this complemented HMS model she does not implicitly know more than she does explicitly. Contrast this with the HMS model to the right. There, she implicitly knows $q$ for instance at state $pq$ (because her implicit possibility set at $pq$ is $\{pq\}$) although she is not aware of $q$ (because her explicit possibility set at $pq$ is on $S_{\{p\}}$). and hence does not explicitly know $q$. The figures demonstrate that the implicit possibility correspondence may be consistent with the explicit possibility correspondence in two different ways. It may model implicit knowledge that is finer than the explicit knowledge (like in the figure to the right) or implicit knowledge that is as coarse as the explicit knowledge but not coarser (like in the figure to the left). Note that a version of the models in Figure~\ref{examples} in which only $\{pq, p \neg q\}$ is in a red dashed oval while $\neg p q$ and $\neg p \neg q$ are in distinct red dashed circles in $S_{pq}$ is ruled out by Projections Preserve Implicit Knowledge.

Given implicit possibility correspondences, we proceed with the definition of the implicit knowledge operators. 
	
\begin{defin}[Implicit Knowledge Operator]\label{definL} For any individual $i \in I$, the \emph{implicit knowledge operator} on events $E \in \Sigma$ is $$L_i(E) := \{\omega \in \Omega : \Lambda_i(\omega) \subseteq E\}$$ if there exists a state $\omega \in \Omega$ such that $\Lambda_i(\omega) \subseteq E$ and by $L_i(E) = \emptyset^{S(E)}$ otherwise. 
\end{defin}
	
The next observation follows immediately from the properties of the implicit possibility correspondence and the proof of Lemma~\ref{alles_events}.
	
\begin{lem}\label{Levent} For any individual $i \in I$ and event $E \in \Sigma$, $L_i(E)$ is an $S(E)$-based event.
\end{lem}

Implicit knowledge satisfies all properties of ``partitional'' knowledge. 
	
\begin{prop}\label{S5_implicit} For any individual $i \in I$, $L_i$ satisfies for any $E, F \in \Sigma$ and $\{E_n\}_n \subseteq \Sigma$,
\begin{itemize}
\item[(i)] Necessitation: For $\Phi \subseteq \mathsf{At}$, $L_i(S_\Phi^\uparrow)= S_{\Phi}^\uparrow$,
			
\item[(ii)] Conjunction: $L_i\left(\bigcap_n E_n\right) = \bigcap_{n} L_i(E_n)$,
			
\item[(iii)] Monotonicity: $E \subseteq F$ implies $L_i(E) \subseteq L_i(F)$,
			
\item[(iv)] Truth: ${L}_i(E) \subseteq E$,
			
\item[(v)] Positive Introspection: $L_i(E) \subseteq L_i L_i(E)$,
			
\item[(vi)] Negative Introspection: $\neg L_i(E) \subseteq L_i \neg L_i(E)$.
\end{itemize}
\end{prop}

\noindent \textsc{Proof.} All properties follow analogously from the proof of Proposition~\ref{strong_explicit_knowledge} in Heifetz, Meier, and Schipper (2006) except for Necessitation and Negative Introspection. Necessitation follows straightforwardly from Strong Confinement. Negative Introspection follows by standard arguments from Stationarity, e.g., Osborne and Rubinstein (1994), p. 70.\hfill $\Box$\\
			
We observe that, as in Fagin and Halpern (1988), explicit knowledge of an event equals to implicit knowledge and awareness of that event.
\begin{prop}\label{prop: semantic equivalence of K - L and A} For any individual $i \in I$ and event $E \in \Sigma$,
	\begin{enumerate}
		\item $K_i(E) = L_i(E) \cap A_i(E)$,
		\item $U_i(E) = L_i(U_i(E))$,
		\item $A_i(E) = L_i(A_i(E))$,
		\item $A_i L_i(E) = A_i(E)$.
	\end{enumerate}
\end{prop}

\noindent  \textsc{Proof.}  1. ``$\subseteq$'': We have $\omega \in K_i(E)$ if and only if $\Pi_i(\omega) \subseteq E$. By Coherence (Lemma~\ref{coherence}) $\Lambda_i(\omega)_{S_{\Pi_i(\omega)}} = \Pi_i(\omega)$. Thus, $\Lambda_i(\omega) \subseteq E$, which is equivalent to $\omega \in L_i(E)$. Moreover, $K_i(E) \subseteq A_i(E)$. Thus, $\omega \in A_i(E)$. Hence, $\omega \in L_i(E) \cap A_i(E)$.
	
``$\supseteq$'': We have $\omega \in A_i(E)$ if and only if $\Pi_i(\omega) \subseteq S \succeq S(E)$. By Coherence (Lemma~\ref{coherence}) of $\Lambda_i$, $\Lambda_i(\omega)_{S_{\Pi_i(\omega)}} \subseteq S \succeq S(E)$. $\omega \in L_i(E)$ if and only if $\Lambda_i(\omega) \subseteq E$. It now follows that $\Lambda_i(\omega)_{S_{\Pi_i(\omega)}} \subseteq E$ and thus by Coherence (Lemma~\ref{coherence}), $\Pi_i(\omega) \subseteq E$, which is equivalent to $\omega \in K_i(E)$.
	
2. ``$\subseteq$'': Let $\w \in U_i(E) = \neg A_i(E)$ and suppose by contradiction that $\w \not\in L_i(U_i(E)) = L_i(\neg A_i(E))$. This means that $\Lambda_i(\w) \not \subseteq \neg A(E)$, i.e., that there exists some $\w' \in \Lambda_i(\w)$ such that $\w' \not\in \neg A(E)$. Since $\w \in \neg A_i(E)$, then $S_{\Pi_i(\w)} \nsucceq S(E)$. However, by Explicit Measurability, since $\w' \in \Lambda_i(\w)$, then $\Pi_i(\w) = \Pi_i(\w')$, which means that $S_{\Pi_i(\w')} \nsucceq S(E)$. This amounts to saying that $\w' \in \neg A(E)$, contradicting the assumption that $\w' \not\in \neg A(E)$. Hence, $\w \in L_i(U_i(E))$.
	
``$\supseteq$'': Immediate by Reflexivity of $\Lambda_i$.

3. ``$\subseteq$'': Let $\w\in A_i(E)$. Then $S_{\Pi_i(w)}\succeq S(E)$. By Explicit Measurability, for all $\w' \in\Lambda_i(\w)$, $\Pi_i(\w) = \Pi_i(\w')$, so for all $\w' \in\Lambda_i(\w)$, $S_{\Pi_i(\w')} \succeq S(E)$. This means that for all $\w' \in \Lambda_i(\w)$, $\w' \in A_i(E)$, i.e., $\Lambda_i(\w) \subseteq A_i(E)$. Thus, $\w \in L_i(A_i(E))$.
	
``$\supseteq$'':  Immediate by Reflexivity of $\Lambda_i$.
	
4. $A_i (L_i(E))= K_i(S(L_i(E))^\uparrow) = K_i(S(E)^\uparrow) = A_i(E)$, where the first and last equality follows from Weak Necessitation and the equality in the middle follows from Lemma~\ref{Levent}. \hfill $\Box$\\

Properties 2. and 3. above mean that the individual implicitly knows her unawareness. This is an aspect where implicit knowledge differs from explicit knowledge. Indeed, by KU introspection (cf. Proposition \ref{prop: K-U properties}), an individual can never explicitly know that she is unaware of an event. Property 4 says that an individual is aware of her implicit knowledge of an event if and only if she is aware of the event. That is, the moment she can reason about an event, she can also reason about her implicit knowledge of the event. This is analogous to AK-Self-Reflection of explicit knowledge.

\section{From Implicit to Explicit Knowledge\label{implicit-to-explicit}}
	
In the previous section, we defined implicit knowledge based on explicit knowledge. In this section, we go the other direction. We can devise a version of HMS models that feature as primitives possibility correspondences capturing implicit knowledge and (non-syntactic) awareness functions, and then derive the possibility correspondence capturing explicit knowledge. 

\begin{defin}[Implicit Knowledge-Based HMS Model] An~\emph{implicit knowledge-based HMS model} $\mathsf{M}^* = \langle I, \{S_{\Phi}\}_{\Phi \subseteq \mathsf{At}}, (r^{\Phi}_{\Psi})_{\Psi \subseteq \Phi \subseteq \mathsf{At}}, (\Lambda^*_i)_{i \in I}, (\alpha_i)_{i \in I}, v \rangle$ consists~of 
\begin{itemize}
\item a non-empty set of individuals $I$,
\item a nonempty collection of nonempty disjoint state spaces $\{S_{\Phi}\}_{\Phi \subseteq \mathsf{At}}$ (as in Definition~\ref{HMS_model}),
\item projections $(r^{\Phi}_{\Psi})_{\Psi \subseteq \Phi \subseteq \mathsf{At}}$ (as in Definition~\ref{HMS_model}),
\item an implicit possibility correspondence $\Lambda_i^*: \Omega \longrightarrow 2^{\Omega} \setminus \{\emptyset\}$, for all $i\in I$,
\item an awareness function $\alpha_i: \Omega \longrightarrow \{S_\Phi\}_{\Phi \subseteq \mathsf{At}}$, for all $i\in I$,
\item a valuation function $v: \mathsf{At} \longrightarrow \Sigma$. 
\end{itemize}
\end{defin}

Like HMS models, implicit knowledge-based HMS models feature a projective lattice of state spaces. However, instead of the explicit possibility correspondence, we now take the implicit possibility correspondences as a primitive. As before, we are interested in strong properties of knowledge associated with S5 because (1) these properties have been used for explicit knowledge in applications, and (2) we will require explicit knowledge to be consistent with implicit knowledge. As such, we are interested in how the rich structure of S5 translates into properties of a derived explicit possibility correspondence. To that end, we require:  

\begin{ass}[Properties of Implicit Possibility Correspondences] For each individual $i \in I$, the implicit possibility correspondence $\Lambda^*_i$ satisfies Reflexivity, Stationarity, and Projections Preserve Implicit Knowledge. 
\end{ass}

These properties were also satisfied by implicit possibility correspondences in the previous section.\footnote{Note again that Reflexivity and Stationarity implies Strong Confinement. In more general settings without Reflexivity or Stationarity, at least Strong Confinement would have to be imposed on $\Lambda_i^*$ for every $i \in I$.}

The second primitive of implicit knowledge-based HMS models is the awareness function $\alpha_i$ for every individual $i \in I$. We impose the following properties on $\alpha_i$: 

\begin{ass}[Properties of Awareness Functions] For each individual $i \in I$, the awareness function $\alpha_i: \Omega \longrightarrow \{S_\Phi\}_{\Phi \subseteq \mathsf{At}}$ satisfies 
\begin{itemize}

\item[O.] Lack of Conception: If $\omega \in S_{\Phi}$, then $\alpha_i(\omega) \preceq S_{\Phi}$.

\item[I.] Awareness Measurability: If $\omega' \in \Lambda^*_i(\omega)$, then $\alpha_i(\omega') = \alpha_i(\omega)$.

\item[II.] If $\omega \in S_{\Phi}$ and $S_{\Psi} \preceq \alpha_i(\omega)$, then $\alpha_i(\omega_{\Psi}) = S_{\Psi}$.

\item[III.] If $\omega \in S_{\Phi}$ and $\alpha_i(\omega) \preceq S_\Psi \preceq S_\Phi$, then $\alpha_i(\omega_{\Psi}) = \alpha_i(\omega)$.

\item[IV.] If $\omega \in S_{\Phi}$ and $\Psi \subseteq \Phi$, then $\alpha_i(\omega) \succeq \alpha_i(\omega_{\Psi})$.
\end{itemize}
When $\alpha_i(\w) \in S$ for some $S \in \{S_{\Phi}\}_{\Phi \subseteq \mathsf{At}}$, we call $S$ the \emph{awareness level of $i$ at $\w$}.
\end{ass}
	
Property O. models one feature of Confinement (cf. Assumption~\ref{assumptions_HMS}). Note that Confinement in HMS models has two features: First, it requires that the possibility set at a state is a subset of exactly one space. Second, it says that this space must be weakly less expressive than the space containing the state. Only this second feature is captured by property O. The idea is that an individual may have lack of conception. Property I. is a measurability condition. Awareness is measurable with respect to implicit knowledge. The implication is that an agent implicitly knows her own awareness. Properties II. to IV. are consistency properties of awareness across the lattice. Projections preserve awareness as long as it is still expressible in the spaces. While property II. preserves awareness for corresponding states in spaces less expressive than the awareness level at a state, property III. preserves awareness for corresponding states in spaces more expressive than the awareness level at that state.

\begin{defin}[Derived Explicit Possibility Correspondence]\label{derived_explicit_possibility_correspondence} Given an implicit \\ knowledge-based HMS model $\mathsf{M}^* = \langle I, \{S_{\Phi}\}_{\Phi \subseteq \mathsf{At}}, (r^{\Phi}_{\Psi})_{\Psi \subseteq \Phi \subseteq \mathsf{At}}, (\Lambda^*_i)_{i \in I}, (\alpha_i)_{i \in I}, v \rangle$, define for each individual $i \in I$ the explicit possibility correspondence $\Pi^*_i: \Omega \longrightarrow 2^{\Omega}$ by, for all $\omega \in \Omega$ and $\Phi \subseteq \mathsf{At}$, $$\Pi^*_i(\omega_{\Phi}) := \Lambda^*_i(\omega)_{\alpha_i(\omega_{\Phi})}.$$ We call \ $\overline{\mathsf{M}}^* = \langle I, \{S_{\Phi}\}_{\Phi \subseteq \mathsf{At}}, (r^{\Phi}_{\Psi})_{\Psi \subseteq \Phi \subseteq \mathsf{At}}, (\Lambda^*_i)_{i \in I}, (\Pi^*_i)_{i \in I}, (\alpha_i)_{i \in I}, v \rangle$ the \emph{complemented implicit knowledge-based HMS model}. 
\end{defin}
	
The defining condition for the explicit possibility correspondence in implicit knowledge-based HMS models is a slight strengthening of Coherence derived from the explicit and implicit measurability in Lemma~\ref{coherence}. Here we take it as the primitive to connect explicit knowledge to implicit knowledge.
	
The following observations are immediate:
\begin{lem}\label{immediate} For all $\omega \in \Omega$,
\begin{itemize}
\item[A.] $\Pi^*_i(\omega) = \Lambda^*_i(\omega)_{\alpha_i(\omega)}$,
\item[B.] $\Pi^*_i(\omega_{\Phi}) = \Lambda^*_i(\omega)_{\Phi}$ for all $\Phi \subseteq \mathsf{At}$ with $S_\Phi \preceq \alpha_i(\omega)$,
\item[C.] $\Pi^*_i(\omega_{\Phi}) = \Lambda^*_i(\omega)_{\alpha_i(\omega)}$ for all $\Phi \subseteq \mathsf{At}$ with $S_{\omega} \succeq S_\Phi \succeq \alpha_i(\omega)$.
\end{itemize}
\end{lem}

\noindent \textsc{Proof.} A. For any $\Psi \subseteq \mathsf{At}$, if $\omega \in S_{\Psi}$, then $\Pi^*_i(\omega) = \Lambda^*_i(\omega)_{\alpha_i(\omega)}$ by definition. 

B. For all $\Phi \subseteq \mathsf{At}$, $\Pi_i^*(\omega_{\Phi}) = \Lambda_i^*(\omega)_{\alpha_i(\omega_{\Phi})}$ by definition. Since $S_{\Phi} \preceq \alpha_i(\omega)$, by II. $\alpha_i(\omega_{\Phi}) = S_{\Phi}$. Thus, $\Pi_i^*(\omega_{\Phi}) = \Lambda_i^*(\omega)_{\Phi}$. 

C. Again, for all $\Phi \subseteq \mathsf{At}$, $\Pi_i^*(\omega_{\Phi}) = \Lambda_i^*(\omega)_{\alpha_i(\omega_{\Phi})}$ by definition. Since in this case $S_{\omega} \succeq S_{\Phi} \succeq \alpha_i(\omega)$, III. implies $\alpha_i(\omega_{\Phi}) = \alpha_i(\omega)$. Hence, $\Pi^*_i(\omega_{\Phi}) = \Lambda_i^*(\omega)_{\alpha_i(\omega)}$.\hfill $\Box$\\

The following sequence of lemmata records properties of the derived explicit possibility correspondence. It shows that it satisfies the properties of the explicit possibility correspondence of HMS models. 
	
\begin{lem}\label{confinement_return} For any individual $i \in I$, if $\alpha_i$ satisfies O., II., and III., then $\Pi^*_i$ satisfies Confinement.
\end{lem}
	
\noindent \textsc{Proof.} If $\omega \in \Omega$, then by A., $\Pi^*_i(\omega) = \Lambda^*_i(\omega)_{\alpha_i(\omega)} \subseteq \alpha_i(\omega)$. By O., $\alpha_i(\omega) \preceq S_\omega$.
	
By C., $\Pi^*_i(\omega_{\Phi}) = \Pi_i(\omega)$ for $S_{\omega} \succeq S_\Phi \succeq \alpha_i(\omega)$. $\Pi^*_i(\omega) \subseteq \alpha_i(\omega)$.
	
By B., $\Pi^*_i(\omega_{\Phi}) = \Lambda^*_i(\omega)_{\Phi}$ for $S_\Phi \preceq \alpha_i(\omega)$. Hence $\Pi^*_i(\omega_{\Phi}) \subseteq S_{\Phi}$. \hfill $\Box$
	
\begin{lem}\label{gr_return} For any individual $i \in I$, if $\alpha_i$ satisfies O., II., and III., then Reflexivity of $\Lambda^*_i$ implies that $\Pi^*_i$ satisfies Generalized Reflexivity.
\end{lem}
	
\noindent \textsc{Proof. } By Reflexivity of $\Lambda^*_i$, $\omega \in \Lambda^*_i(\omega)$. By O., $\Lambda^*_i(\omega) \subseteq (\Lambda^*_i(\omega)_{\alpha_i(\omega)})^{\uparrow}$. Thus, by A. $\omega \in (\Lambda^*_i(\omega)_{\alpha_i(\omega)})^{\uparrow} = \Pi_i^{* \uparrow}(\omega)$. For any $S_\Phi \preceq \alpha_i(\omega)$, $\omega_{\Phi} \in \Lambda^*_i(\omega)_\Phi = \Pi^*_i(\omega_{\Phi})$ by B. For any $\Phi \subseteq \mathsf{At}$ with $S_\omega \succeq S_\Phi \succeq \alpha_i(\omega)$, $\omega_{\Phi} \in (\Lambda^*_i(\omega)_{\alpha_i(\omega)})^{\uparrow} = \Pi^{* \uparrow}_i(\omega_{\Phi})$ by C.\hfill $\Box$
	
\begin{lem}\label{stationarity_return} For any individual $i \in I$, if $\alpha_i$ satisfies I., II., and III., then Stationarity of $\Lambda^*_i$ implies that $\Pi^*_i$ satisfies Stationarity.
\end{lem}
	
\noindent \textsc{Proof.} For any $\omega \in \Omega$, A. implies $\Lambda^*_i(\omega)_{\alpha_i(\omega)} = \Pi^*_i(\omega)$. So we have $\omega' \in \Pi^*_i(\omega)$ if and only if $\omega' \in \Lambda^*_i(\omega)_{\alpha_i(\omega)}$. Then we know that there exists $\omega'' \in (r^{S_{\omega}}_{\alpha_i(\omega)})^{-1}(\{\omega'\})$ with $\omega'' \in \Lambda^*_i(\omega)$. By Stationarity of $\Lambda^*_i$, $\Lambda^*_i(\omega'') = \Lambda^*_i(\omega)$. Thus, by I. $\Lambda^*_i(\omega'')_{\alpha_i(\omega'')} = \Lambda^*_i(\omega)_{\alpha_i(\omega)}$ and hence by A., $\Pi^*_i(\omega'') = \Pi^*_i(\omega)$. Since $\omega''_{\alpha_i(\omega'')} = \omega'$, by B. $\Pi^*_i(\omega') = \Pi^*_i(\omega)$.
	
Now consider the case $S_\Phi \preceq \alpha_i(\omega)$ and $\omega' \in \Pi^*_i(\omega_{\Phi})$. The arguments are analogous to previous arguments except that from I. we conclude $\Lambda^*_i(\omega'')_{\Phi} = \Lambda^*_i(\omega)_{\Phi}$.
	
Finally, consider the case $\Phi \subseteq \mathsf{At}$ with $S_{\omega} \succeq S_\Phi \succeq \alpha_i(\omega)$ and $\omega' \in \Pi^*_i(\omega_{\Phi})$. The arguments are analogous to the arguments in the first paragraph of the proof except that $\omega$ is replaced by $\omega_{\Phi}$. \hfill $\Box$
	
\begin{lem}\label{PPI_return} For any individual $i \in I$, if $\alpha_i$ satisfies II., III., and IV., then $\Pi^*_i$ satisfies Projections Preserve Ignorance.
\end{lem}
	
\noindent \textsc{Proof.} For any $\omega \in \Omega$, A. implies $\Lambda_i^{* \uparrow}(\omega)_{\alpha_i(\omega)} = \Pi_i^{* \uparrow}(\omega)$.
	
If $S_\Psi \preceq \alpha_i(\omega)$, then B. implies $\Pi_i^{* \uparrow}(\omega) \subseteq \Lambda_i^{* \uparrow}(\omega)_{\Psi} = \Pi_i^{* \uparrow}(\omega_{\Psi})$.
	
If $\Psi \subseteq \mathsf{At}$ with $S_{\omega} \succeq S_\Psi \succeq \alpha_i(\omega)$, then C. implies $\Pi_i^{* \uparrow}(\omega) = \Pi_i^{* \uparrow}(\omega_{\Psi})$.
	
If $\Psi \subseteq \mathsf{At}$ such that $S_{\omega} \succeq S_{\Psi}$ and $S_\Psi$ is incomparable to $\alpha_i(\omega)$, then $\Lambda_i^{* \uparrow}(\omega)_{\alpha_i(\omega)} \subseteq \Lambda_i^{* \uparrow}(\omega)_{\alpha_i(\omega_{\Psi})}$ since by IV., $\alpha_i(\omega) \succeq \alpha_i(\omega_{\Psi})$. Thus, $\Pi_i^{* \uparrow}(\omega) \subseteq \Pi_i^{* \uparrow}(\omega_{\Psi})$. \hfill $\Box$
	
\begin{lem}\label{PPK_return} For any individual $i \in I$, if $\alpha_i$ satisfies II. and III., then $\Pi^*_i$ satisfies Projections Preserve Knowledge.
\end{lem}
	
\noindent \textsc{Proof.} For any $\omega \in \Omega$, $\Lambda^*_i(\omega)_{\alpha_i(\omega)} = \Pi^*_i(\omega) \subseteq \alpha_i(\omega)$ by A. 

For $\Phi \subseteq \mathsf{At}$ with $S_{\Phi} \preceq \alpha_i(\omega)$, $\Pi^*_i(\omega_{\Phi}) = \Lambda^*_i(\omega)_{\Phi}$ by B. It follows that $\Pi^*_i(\omega)_{\Phi} = \Pi^*_i(\omega_{\Phi})$. Hence, $\Pi^*_i(\omega)_{\Upsilon} = \Pi^*_i(\omega_{\Upsilon})$ for $\Upsilon \subseteq \Phi$.
	
For $\Phi \subseteq \mathsf{At}$ with $S_{\omega} \succeq S_{\Phi} \succeq \alpha_i(\omega)$, $\Pi^*_i(\omega_{\Phi}) = \Lambda^*_i(\omega)_{\alpha_i(\omega)}$ by C. Hence, $\Pi_i^*(\omega_{\Phi}) = \Pi_i^*(\omega)$. It follows that $\Pi^*_i(\omega_{\Phi})_{\Upsilon} = \Pi_i^*(\omega)_{\Upsilon} = \Pi^*_i(\omega_{\Upsilon})$ for $S_\Upsilon \preceq \alpha_i(\omega)$. \hfill $\Box$\\
	
We conclude that the derived explicit possibility correspondence $\Pi^*_i$ is a possibility correspondence as in Heifetz, Meier, and Schipper (2006, 2008), i.e., satisfies Assumption~\ref{assumptions_HMS}. To show that the connection between the derived explicit possibility correspondence and the implicit possibility correspondence is as in the complemented HMS model of the prior section, we note the following lemma. 

\begin{lem}\label{measurability_return} For any individual $i \in I$, $\Lambda_i^*$ and $\Pi_i^*$ jointly satisfy explicit and implicit measurability. 
\end{lem}

\noindent \textsc{Proof. } For explicit measurability, we need to show that for all $\omega \in \Omega$, $\omega' \in \Lambda_i^*(\omega)$ implies $\Pi_i^*(\omega') = \Pi_i(\omega)$. By Stationarity of $\Lambda_i^*$, $\omega' \in \Lambda^*_i(\omega)$ implies $\Lambda_i^*(\omega') = \Lambda^*(\omega)$. By Lemma~\ref{immediate} A., we have $\Pi^*_i(\omega') = \Lambda_i^*(\omega')_{\alpha_i(\omega')}$ and $\Pi^*_i(\omega) = \Lambda_i^*(\omega)_{\alpha_i(\omega)}$. By Awareness Measurability, $\omega' \in \Lambda_i^*(\omega)$ implies $\alpha_i(\omega') = \alpha_i(\omega)$. Thus, $\Pi_i^*(\omega') = \Pi_i(\omega)$.

For implicit measurability, we need to show that for all $\omega \in \Omega$, $\omega' \in \Pi_i^*(\omega)$, $\Lambda_i^*(\omega') = \Lambda_i^*(\omega)_{S_{\Pi_i^*(\omega)}}$. By Lemma~\ref{stationarity_return} (Stationarity), $\omega' \in \Pi_i^*(\omega)$, $\Pi_i^*(\omega') = \Pi_i^*(\omega)$. This implies that $S_{\omega'} = S_{\Pi_i^*(\omega)}$. Since by Lemma~\ref{immediate} A., we have $\Pi^*_i(\omega') = \Lambda_i^*(\omega')_{\alpha_i(\omega')}$ and $\Pi^*_i(\omega) = \Lambda_i^*(\omega)_{\alpha_i(\omega)}$, it follows now $\alpha_i(\omega') = \alpha_i(\omega) = S_{\Pi^*_i(\omega)}$. From Strong Confinement of $\Lambda_i^*$ (implied by Reflexivity and Stationarity of $\Lambda_i^*$) follows that $\omega' \in \Lambda_i^*(\omega')$. Hence, $\Lambda_i^*(\omega') \subseteq S_{\Pi_i(\omega)}$. It follows now that $\Lambda_i^*(\omega') = \Pi^*_i(\omega') = \Pi_i^*(\omega) = \Lambda^*_i(\omega)_{S_{\Pi_i(\omega)}}$. \hfill $\Box$\\

The above lemmata imply the following: 

\begin{cor}\label{cHMS_derived} For any implicit knowledge-based HMS model $\mathsf{M}^* = \langle I, \{S_{\Phi}\}_{\Phi \subseteq \mathsf{At}}, \\ (r^{\Phi}_{\Psi})_{\Psi \subseteq \Phi \subseteq \mathsf{At}}, (\Lambda^*_i)_{i \in I}, (\alpha_i)_{i \in I}, v \rangle$ with derived explicit possibility correspondences $(\Pi^*_i)_{i \in I}$ we have that $\overline{\mathsf{M}} = \langle I, \{S_{\Phi}\}_{\Phi \subseteq \mathsf{At}}, (r^{\Phi}_{\Psi})_{\Psi \subseteq \Phi \subseteq \mathsf{At}}, (\Lambda^*_i)_{i \in I}, (\Pi^*_i)_{i \in I}, v \rangle$ is a complemented HMS model and $\mathsf{M} = \langle I, \{S_{\Phi}\}_{\Phi \subseteq \mathsf{At}}, (r^{\Phi}_{\Psi})_{\Psi \subseteq \Phi \subseteq \mathsf{At}}, (\Pi^*_i)_{i \in I}, v \rangle$ is a HMS model. 
\end{cor}

The awareness function can be directly used to define an awareness operator on events. 

\begin{defin}[Awareness Operator II] For each individual $i \in I$, define an awareness operator on events by for all $E \in \Sigma$, $$A^*_i(E) := \{\omega \in \Omega : \alpha_i(\omega) \succeq S(E)\}$$ if there is a state $\omega \in \Omega$ such that $\alpha_i(\omega) \succeq S(E)$ and by $A^*_i(E) = \emptyset^{S(E)}$ otherwise.
\end{defin} 

Similarly, for each individual $i \in I$, we can use the possibility correspondence $\Lambda^*_i$ to define an implicit knowledge operator $L^*_i$ as in Definition~\ref{definL}. Finally, let $K_i$ be the explicit knowledge operator and $A_i$ be the awareness operator defined from the derived explicit possibility correspondence $\Pi^*_i$ as in Definitions~\ref{definK} and~\ref{definA}, respectively.
 
The following proposition shows that awareness defined with the awareness function is equivalent to awareness defined with the derived explicit possibility correspondence. It also shows that explicit knowledge defined from the derived explicit possibility correspondence is equivalent to implicit knowledge and awareness defined from the primitive implicit possibility correspondence and awareness function.

\begin{prop}\label{otherwayaround} For every individual $i \in I$ and any event $E \in \Sigma$,
\begin{enumerate}
\item $A^*_i(E) = A_i(E)$
\item $K_i(E) = L^*_i(E) \cap A^*_i(E)$.
\end{enumerate}
\end{prop}
	
\noindent \textsc{Proof.} 1. For any $E \in \Sigma$, $\omega \in A^*_i(E)$ if and only if $\alpha_i(\omega) \succeq S(E)$. By definition of $\Pi^*_i$, this holds if and only if $\Pi^*_i(\omega) \subseteq S \succeq S(E)$. This is equivalent to $\omega \in A_i(E)$. 
	
2. For any non-empty $E \in \Sigma$, $\omega \in K_i(E)$ if and only if $\Pi^*_i(\omega) \subseteq E$. By definition of $\Pi^*_i$, it holds if and only if $\Lambda^*_i(\omega) \subseteq E$ and $\alpha_i(\omega) \succeq S(E)$. This is equivalent to $\omega \in L^*_i(E)$ and $\omega \in A^*_i(E)$. If $E = \emptyset^{S_{\Phi}}$ for some $\Phi \subseteq \mathsf{At}$, then $K_i(E) = \emptyset^{S_{\Phi}}$ and $L^*_i(E) = \emptyset^{S_{\Phi}}$. $A^*_i(E) \subseteq S_{\Phi}^{\uparrow}$. Thus, by conjunction of events $L^*_i(E) \cap A^*_i(E) = \emptyset^{S_{\Phi}}$. \hfill $\Box$\\

Together, Sections~\ref{explicit-to-implicit} and~\ref{implicit-to-explicit} show an interdefinability of explicit and implicit knowledge in HMS models. Implicit knowledge can be defined in terms of explicit knowledge and explicit knowledge can be defined in terms of implicit knowledge. We can use either the explicit possibility correspondences as primitive or the implicit possibility correspondence together with the awareness function. Implicit knowledge-based HMS models are arguably closer to FH models than HMS models. We will use them to build a bridge between HMS and FH models.

\section{Category of FH Models}
	
In this section, we introduce FH models and bounded morphisms, a notion of structure preserving maps between FH models, and use these to form a category with FH models as objects and bounded morphisms as morphisms. As mentioned in the introduction, the goal is to provide subjective versions of FH models. 

The semantics of FH models is not syntax-free since each agent's awareness function assigns to each state a set of formulas. Thus, we first introduce the formal language featuring implicit knowledge, awareness, and explicit knowledge. With $i \in I$ and $p \in \mathsf{At}$, define the language $\mathcal{L}_{\mathsf{At}}$ by
$$\varphi::=\top \mid p \mid \neg\varphi \mid \varphi \wedge \psi \mid \ell_i \varphi \mid a_i \varphi \mid k_i \varphi$$
	
With some abuse of notation, let $\mathsf{At}(\varphi) := \{ p \in \mathsf{At} \colon p \text{ is a subformula of } \varphi\}$, for any $\varphi \in \mathcal{L}_{\mathsf{At}}$, be the set of atomic formulas that are contained in $\varphi$, and let $\mathcal{L}_\Phi := \{\varphi \in \mathcal{L}_{\mathsf{At}}: \mathsf{At}(\varphi) \subseteq \Phi \}$ be the sublanguage of $\mathcal{L}_{\mathsf{At}}$ built on propositions $p$ in $\Phi \subseteq \mathsf{At}$. 
	
The formula $\ell_i \varphi$ reads ``agent $i$ implicitly knows $\varphi$",  $a_i \varphi$ reads ``$i$ is aware of $\varphi$", and $k_i \varphi$ reads ``$i$ explicitly knows $\varphi$". Fagin and Halpern (1988) define explicit knowledge as the conjunction of implicit knowledge and awareness, namely $k_{i} \varphi = a_i \varphi \wedge \ell_i \varphi$, for $\varphi \in \mathcal{L}_{\mathsf{At}}$.

\begin{defin}[FH Model]\label{def:FH model} For any $\Phi \subseteq \mathsf{At}$, an \emph{FH model} $\mathsf{K}_{\Phi} = \langle I, W_{\Phi}, (R_{\Phi, i})_{i \in I}, \\ (\mathcal{A}_{\Phi,i})_{i \in I}, V_{\Phi}\rangle$ for $\Phi$ consists of 
\begin{itemize}
\item a non-empty set of individuals $I$,
\item a non-empty set of states $W_{\Phi}$,
\item an accessibility relation $R_{\Phi, i} \subseteq W_{\Phi} \times W_{\Phi}$, for all $i\in I$,
\item an awareness function $\mathcal{A}_{\Phi, i}: W_{\Phi} \longrightarrow 2^{\mathcal{L}_{\Phi}}$, for all $i\in I$, assigning to each state $w \in W_{\Phi}$ a set of formulas $\mathcal{A}_{\Phi, i}(w) \subseteq \mathcal{L}_{\Phi}$. The set $\mathcal{A}_{\Phi, i}(w)$ is called the \emph{awareness set of $i$ at $w$}. 
\item a valuation function $V_{\Phi}: \Phi \longrightarrow 2^{W_{\Phi}}$. 
\end{itemize}
\end{defin} 

\begin{ass}[Properties Imposed on FH Models] We require that the FH model $K_{\Phi}$ is \emph{propositionally determined}, i.e., for every $i \in I$, the awareness functions satisfy\footnote{The terminology is from Halpern (2001).}
\begin{itemize}
\item[] \emph{Awareness is Generated by Primitive Propositions:} For all $\varphi \in \mathcal{L}_{\Phi}$, $\varphi \in \mathcal{A}_{\Phi, i}(w)$ if and only if for all $p \in \mathsf{At}(\varphi)$, $p \in \mathcal{A}_{\Phi, i}(w)$.
\item[] \emph{Agents Know What They Are Aware of:} $(w, t) \in R_{\Phi, i}$ implies $\mathcal{A}_{\Phi, i}(w) = \mathcal{A}_{\Phi, i}(t)$.
\end{itemize}
We also require that the FH model $\mathsf{K}_{\Phi}$ is \emph{partitional}, that is, $R_{\Phi, i}$ is an equivalence relation, i.e., satisfies reflexivity, transitivity, and Euclideaness. 
\end{ass} Throughout the paper, we focus on partitional and propositionally determined FH models because these models capture the notion of awareness and knowledge used in most applications so far and it is also the notion of awareness used in HMS models. We are interested in how this rich structure maps between FH models as well as between FH and HMS models. 
 
\begin{defin}[Bounded Morphism] For any $\Psi \subseteq \Phi \subseteq \mathsf{At}$ and FH models $\mathsf{K}_{\Phi} = \langle I, W_{\Phi}, (R_{\Phi, i})_{i \in I}, (\mathcal{A}_{\Phi,i})_{i \in I}, V_{\Phi}\rangle$ and $\mathsf{K}_{\Psi} = \langle I, W_{\Psi}, (R_{\Psi, i})_{i \in I}, (\mathcal{A}_{\Psi,i})_{i \in I}, V_{\Psi}\rangle$, the mapping $f^{\Phi}_{\Psi}: \mathsf{K}_{\Phi} \longrightarrow \mathsf{K}_{\Psi}$ is a \emph{surjective bounded morphism} if for every $i \in I$ and $w\in W_\Phi$
\begin{itemize}
\item \emph{Surjectivity:} $f^{\Phi}_{\Psi}: W_{\Phi} \longrightarrow W_{\Psi}$ is a surjection,

\item \emph{Atomic harmony:} for every $p \in \Psi$, $w \in V_{\Phi}(p)$ if and only if $f^{\Phi}_{\Psi}(w) \in V_{\Psi}(p)$,

\item \emph{Awareness consistency:} $\mathcal{A}_{\Phi, i}(w) \cap \mathcal{L}_{\Psi} = \mathcal{A}_{\Psi, i}(f^{\Phi}_{\Psi}(w))$,

\item \emph{Homomorphism:} $f^{\Phi}_{\Psi}$ is a homomorphism with respect to $R_{\Phi, i}$, i.e., if $(w, t) \in R_{\Phi, i}$, then $(f^{\Phi}_{\Psi}(w),f^{\Phi}_{\Psi}(t))\in R_{\Psi,i}$, 
\item \emph{Back:}  
if $(f^{\Phi}_{\Psi}(w),t') \in R_{\Psi, i}$, then there is a state $t \in W_{\Phi}$ such that $f^{\Phi}_{\Psi}(t) = t'$ and $(w, t) \in R_{\Phi, i}$.
\end{itemize}
\end{defin}

This is the standard notion of surjective bounded morphism (also called surjective $p$-morphism) (see for instance, Blackburn, de Rijke, and Venema, 2001, pp. 59--62) except for the additional property of Awareness Consistency. In our context, the bounded morphism literally bounds the language over which FH models are defined. We can now consider collections of FH models and commuting bounded morphisms between them as follows: 
 
\begin{defin}[Category of FH Models]\label{category} Given the FH model $\mathsf{K}_{\mathsf{At}}$, the \emph{category of FH models} $\mathcal{C}(\mathsf{K}_{\mathsf{At}}) = \langle (\mathsf{K}_{\Phi})_{\Phi \subseteq \mathsf{At}}, (f^{\Phi}_{\Psi})_{\Psi \subseteq \Phi \subseteq \mathsf{At}} \rangle$ consists of 
\begin{itemize}
\item a collection of FH models $\mathsf{K}_{\Phi}$, one for each $\Phi \subseteq \mathsf{At}$,
\item for any $\Phi, \Psi \subseteq \mathsf{At}$ with $\Psi \subseteq \Phi$, there is one surjective bounded morphism $f^{\Phi}_{\Psi}$, such that 

\smallskip
-- for any $\Phi \subseteq \mathsf{At}$, $f^{\Phi}_{\Phi}$ is the identity,
 
\smallskip
-- for any $\Upsilon, \Phi, \Psi \subseteq \mathsf{At}$ with $\Upsilon \subseteq \Psi \subseteq \Phi$, $f^{\Phi}_{\Upsilon} = f^{\Psi}_{\Upsilon} \circ f^{\Phi}_{\Psi}$. 
\end{itemize}
 
\end{defin}

Our terminology is not arbitrary. The category of FH models is indeed a category in the sense of category theory. It has an initial object, the most expressive FH model $\mathsf{K}_{\mathsf{At}}$, as well as a terminal object, the FH model $\mathsf{K}_{\emptyset}$. 

Since the category of FH models is defined with bounded morphisms, it suggests that all FH models in the category are in some sense epistemically equivalent. Indeed, we interpret each category of FH models literally as the category of FH models that vary with the language but are otherwise modally equivalent. That is, for any $\Psi \subseteq \Phi \subseteq \mathsf{At}$, modal satisfaction for $\mathsf{K}_{\Psi}$ is as for $\mathsf{K}_{\Phi}$ with respect to formulas in $\mathcal{L}_{\Psi}$ (see Lemma~\ref{boring_modal_equivalence} below). We interpret this as follows: When a modeler represents a context with a FH model $\mathsf{K}_{\mathsf{At}}$, an agent $i$ at state $w \in W_{\mathsf{At}}$ can be thought of representing it with the FH model $\mathsf{K}_{\mathsf{At}(\mathcal{A}_{\mathsf{At}, i}(w))}$.\footnote{As usual, $\mathsf{At}(\mathcal{A}_{\mathsf{At}, i}(w))$ denotes the image under the function $\mathsf{At}(\cdot)$ of $\mathcal{A}_{\mathsf{At}, i}(w)$.} And this agent $i$ considers it possible at $w$ that at $t$ with $(f^{\mathsf{At}}_{\mathsf{At}(\mathcal{A}_{\mathsf{At}, i}(w))}(w), t) \in R_{\mathsf{At}(\mathcal{A}_{\mathsf{At}, i}(w)), i}$ agent $j$ represents the situation with the FH model $\mathsf{K}_{\mathsf{At}(\mathcal{A}_{\mathsf{At}, j}(t))}$, etc. These models can all be seen as equivalent except for the language of which they are defined. With this construction, we do not just endow agents with a formal language to reason about their context but we also allow them to analyze their context with semantic devices like logicians do. This is relevant because in many multi-agent contexts of game theory, the analysis proceeds using semantic devices like state spaces etc. rather than at the level of syntax. For instance, in a principal-agent problem, the principal may want to use a FH model augmented by actions and utility functions to analyze optimal contract design realizing that the (unaware) agent may also use a less expressive but otherwise equivalent FH model to analyze how to optimally interact with the principal. 

To make the equivalence between models in the category of FH models precise, we need to introduce the semantics of FH models. 
\begin{defin}[FH Semantics] For any $\Phi \subseteq \mathsf{At}$, FH model $\mathsf{K}_{\Phi} = \langle I, W_{\Phi}, (R_{\Phi, i})_{i \in I}, (\mathcal{A}_{\Phi,i})_{i \in I}, V_{\Phi}\rangle$, and $\omega \in W_{\Phi}$, satisfaction of formulas in $\mathcal{L}_{\Phi}$ is given by the following clauses:
\begin{center}
	\begin{tabular}{lllcclll}
		    $\mathsf{K}_{\Phi}, w \Vdash \top$ &  & for all $w\in W_\Phi$; & & & $\mathsf{K}_{\Phi}, w \Vdash \varphi\wedge \psi$ &  iff & $\mathsf{K}_{\Phi}, w \Vdash \varphi$ and $\mathsf{K}_{\Phi}, w \Vdash \psi$; \tabularnewline
			$\mathsf{K}_{\Phi}, w \Vdash p$ &  iff & $w \in V_{\Phi}(p)$; & & & $\mathsf{K}_{\Phi},w \Vdash a_i\varphi$ & iff & $\varphi\in\mathcal{A}_{\Phi, i}(w)$; \tabularnewline
			$\mathsf{K}_{\Phi}, w \Vdash \neg \varphi$ &  iff & $\mathsf{K}_{\Phi}, w \not\Vdash \varphi$; & & & $\mathsf{K}_{\Phi}, w \Vdash \ell_i \varphi$ & iff & $\mathsf{K}_{\Phi}, t\Vdash \varphi$ for all $(w, t) \in R_{\Phi, i}$.\tabularnewline		
	\end{tabular}
\end{center}
\end{defin}
	
From this semantics and the syntactic definition $k_i\varphi := \ell_i \varphi \wedge a_i \varphi$, it follows that $\mathsf{K}_{\Phi}, w \Vdash k_i \varphi$ if and only if for all $t$ such that $(w, t) \in R_{\Phi, i}$, $\mathsf{K}_{\Phi}, t \Vdash \varphi$ and $\varphi \in \mathcal{A}_{\Phi, i}(w)$.

The category of FH models forms a complete lattice induced by set inclusion on sets of atomic formulas with the initial object being the meet of the lattice and the terminal object being the join of the lattice. We now show that it gives rise to a complete lattice when ordered using the (directed) bounded morphism or, epistemically more relevant, when ordered by modal equivalence relative to sublanguages. To prove the proposition showing that a category forms a complete lattice, we need the following lemma, which shows that for any two FH models in the category, they are modally equivalent with respect to the language for which the less expressive FH models is defined.

\begin{lem}\label{boring_modal_equivalence} Given a category of FH models, $\langle (\mathsf{K}_{\Phi})_{\Phi \subseteq \mathsf{At}}, (f^{\Phi}_{\Psi})_{\Psi \subseteq \Phi \subseteq \mathsf{At}} \rangle$, for any $\Psi, \Phi \subseteq \mathsf{At}$ with $\Psi \subseteq \Phi$, all $w \in W_{\Phi}$, and all $\varphi \in \mathcal{L}_{\Psi}$, $$\mathsf{K}_{\Phi}, w \Vdash \varphi \mbox{ if and only if } \mathsf{K}_{\Psi}, f^{\Phi}_{\Psi}(w) \Vdash \varphi.$$
\end{lem}

\noindent \textsc{Proof. } The proof is by induction on the complexity of formulas in $\mathcal{L}_{\Psi}$. The base case follows directly from Atomic harmony. The Boolean cases are straightforward once we note that these formulas are restricted to $\mathcal{L}_{\Psi}$. 

Now consider the case $\ell_i \varphi \in \mathcal{L}_{\Psi}$:

``$\Rightarrow$'': Assume $\mathsf{K}_{\Phi}, w \Vdash \ell_i \varphi$. This holds if and only if $\mathsf{K}_{\Phi}, w' \Vdash \varphi$ for all $(w, w') \in R_{\Phi, i}$. By the induction hypothesis, $\mathsf{K}_{\Psi}, f^{\Phi}_{\Psi}(w') \Vdash \varphi$. Since $f^{\Phi}_{\Psi}$ is a homomorphism, $(f^{\Phi}_{\Psi}(w), f^{\Phi}_{\Psi}(w')) \in R_{\Psi, i}$. Suppose by contradiction that there exists $w'' \in W_{\Psi}$ such that $\mathsf{K}_{\Psi}, w'' \not\Vdash \varphi$ but $(f^{\Phi}_{\Psi}(w), w'') \in R_{\Psi, i}$. By the Back condition, there exists $w''' \in W_{\Phi}$ such that $f^{\Phi}_{\Psi}(w''') = w''$ and $(w, w''') \in R_{\Phi, i}$. By Atomic harmony, $\mathsf{K}_{\Phi}, w''' \not\Vdash \varphi$. This contradicts now $\mathsf{K}_{\Phi}, w' \Vdash \varphi$ for all $(w, w') \in R_{\Phi, i}$. 

``$\Leftarrow$'': Assume $\mathsf{K}_{\Psi}, f^{\Phi}_{\Psi}(w) \Vdash \ell_i \varphi$. This holds if and only if $\mathsf{K}_{\Psi}, w' \Vdash \varphi$ for all $(f^{\Phi}_{\Psi}(w), w') \in R_{\Psi, i}$. Suppose by contradiction that there exists $w'' \in (f^{\Phi}_{\Psi})^{-1}(w)$ such that $\mathsf{K}_{\Phi}, w''' \not\Vdash \varphi$ for some $w''' \in W_{\Phi}$ with $(w'', w''') \in R_{\Phi, i}$. By the induction hypothesis, $\mathsf{K}_{\Psi}, f^{\Phi}_{\Psi}(w''') \not\Vdash \varphi$. Since $f^{\Phi}_{\Psi}$ is a homomorphism, $(f^{\Phi}_{\Psi}(w''), f^{\Phi}_{\Psi}(w''')) \in R_{\Psi, i}$. But $f^{\Phi}_{\Psi}(w'') = f^{\Phi}_{\Psi}(w)$. Thus, we have a contradiction to $\mathsf{K}_{\Psi}, w' \Vdash \varphi$ for all $(f^{\Phi}_{\Psi}(w), w') \in R_{\Psi, i}$.

Now consider the case $a_i \varphi \in \mathcal{L}_{\Psi}$: 

We have that $\mathsf{K}_{\Phi}, w \Vdash a_i \varphi$ if and only if $\varphi \in \mathcal{A}_{\Phi, i}(w)$ by the definition of semantics. By Awareness consistency, $\mathcal{A}_{\Psi, i}(f^{\Phi}_{\Psi}(w)) = \mathcal{A}_{\Phi, i}(w) \cap \mathcal{L}_{\Psi}$. Moreover, $a_i \varphi \in \mathcal{L}_{\Psi}$ implies $\varphi \in \mathcal{L}_{\Psi}$. Thus, $\varphi \in \mathcal{A}_{\Psi, i}(f^{\Phi}_{\Psi}(w))$. This holds if and only if $\mathsf{K}_{\Psi}, f^{\Phi}_{\Psi}(w) \Vdash a_i \varphi$ (by the definition of the semantics).\hfill $\Box$\\

We are now ready to claim that the category of FH models is a complete lattice when ordered by modal equivalence modulo sublanguages. 

\begin{prop}\label{epistemic_equivalence} Given a category of FH models, $\langle (\mathsf{K}_{\Phi})_{\Phi \subseteq \mathsf{At}}, (f^{\Phi}_{\Psi})_{\Psi \subseteq \Phi \subseteq \mathsf{At}} \rangle$, modal equivalence relative to sublanguages forms a complete lattice of FH models in the category as follows: For any nonempty set of subsets of atomic formulas $\mathcal{F} \subseteq 2^{\mathsf{At}}$,   
\begin{itemize}
\item[(i)] $\mathsf{K}_{\bigcup_{\Phi \in \mathcal{F} } \Phi}$ is modally equivalent to $\mathsf{K}_{\Psi}$ w.r.t. $\mathcal{L}_{\Psi}$ for every $\Psi \in \mathcal{F} $, i.e., for any $w \in W_{\bigcup_{\Phi \in \mathcal{F} } \Phi}$, $\varphi \in\mathcal{L}_\Psi$, $\mathsf{K}_{\bigcup_{\Phi \in \mathcal{F} } \Phi}, w \Vdash \varphi$ iff $\mathsf{K}_\Psi, f^{\bigcup_{\Phi \in \mathcal{F} } \Phi}_{\Psi}(w) \Vdash \varphi$, and 
\item[(ii)] $\mathsf{K}_{\bigcap_{\Phi \in \mathcal{F}} \Phi}$ is modally equivalent to $\mathsf{K}_{\Psi}$ w.r.t. $\mathcal{L}_{\bigcap_{\Phi \in \mathcal{F}}}$ for every $\Psi \in \mathcal{F}$, i.e., for any $w \in W_{\Psi}$, $\varphi \in \mathcal{L}_{\bigcap_{\Phi \in \mathcal{F}} \Phi}$, $\mathsf{K}_\Psi, w \Vdash \varphi$ iff $\mathsf{K}_{\bigcap_{\Phi \in \mathcal{F}} \Phi}, f^{\Psi}_{\bigcap_{\Phi \in \mathcal{F} } \Phi}(w) \Vdash \varphi$.
\end{itemize}
\end{prop}

\noindent \textsc{Proof.} By Lemma~\ref{boring_modal_equivalence}, modal equivalence w.r.t. sublanguages is an order on objects in the category of FH models. To show that modal equivalence w.r.t. sublanguages forms a complete lattice of FH models, notice that reflexivity follows from the requirement of the FH category that the bounded morphism from the FH model to itself is the identity. Antisymmetry follows from the requirement of the FH category that if there is a bounded morphism from one FH model to another and back, then it must be identity. Transitivity follows from composition of bounded morphisms. To show that for every subset of FH models there is a join and a meet, we need to prove (i) and (ii) respectively. To prove (i), let $\Phi = \bigcup_{\Upsilon \in \mathcal{F}} \Upsilon$ and $\Psi \in \mathcal{F}$ and apply Lemma~\ref{boring_modal_equivalence}. To prove (ii), let $\Psi = \bigcap_{\Upsilon \in \mathcal{F}} \Upsilon$ and $\Phi \in \mathcal{F}$ and apply Lemma~\ref{boring_modal_equivalence}. This completes the proof of the proposition. \hfill $\Box$\\

Note that for a collection of FH models $\{\mathsf{K}_{\Psi}\}_{\Psi \in \mathcal{F}}$, the ``join'' and ``meet'' FH models are $\mathsf{K}_{\bigcup_{\Phi \in \mathcal{F} } \Phi}$ and $\mathsf{K}_{\bigcap_{\Phi \in \mathcal{F}} \Phi}$, respectively. So for any collection of FH models, Proposition~\ref{epistemic_equivalence} shows modal equivalence between any FH model in the collection and its join and meet models, respectively.

Our notion of bounded morphism  is inspired by bisimulation of FH models introduced by van Ditmarsch et al. (2018). Clearly, the surjective bounded morphism is a bisimulation (Blackburn et al., 2001). Here we discuss some differences and similarities between the notions. While bisimulation more generally is a relation between models without a particular direction, the bounded morphism has a natural direction from the more expressive FH model to the less expressive FH model. Further, it is a function on $W_{\Phi}$. That is, it maps \emph{every} state in $W_{\Phi}$ to a state in $W_{\Psi}$ with $\Psi \subseteq \Phi$. Moreover, surjectivity is a property that is straightforward to define for functions. Finally, bounded morphisms easily compose and almost naturally lead to the notion of category of FH models although we do not really make much use here of the machinery of category theory.\footnote{There is a growing literature making use of categories of Kripke frames with bounded morphisms between them, e.g., Moss (1999).} For all these reasons, we use the notion of bounded morphism. Van Ditmarsch et al. (2018) introduced two notions of bisimulation for FH models, standard bisimulation and awareness bisimulation. Like our bounded morphism, both of their notions of bisimulation also depend on a subset of atomic formulas for FH models. The clauses Atomic harmony, Awareness consistency, Homomorphism, and Back have counterparts in their notions of bisimulations. Our notion of bounded morphism is closer to what they call standard bisimulation because our Homomorphism and Back clauses do not involve the awareness function. Although van Ditmarsch et al. (2018, p. 63) mention the projective lattice structure of Heifetz, Meier, and Schipper (2006, 2008) as a motivation for their notion of awareness bisimulation, we believe it is particularly useful for their notion of speculative knowledge. Their notions of bisimulations do not explicitly require surjectivity. However, when they consider maximal bisimulations, they must be automatically surjective since they yield quotient models. Compositions of maximal bisimulation commute like we require our bounded morphism to do in the category of FH models but bisimulations that are not maximal do not necessarily commute. Moreover, maximal bisimulations yield necessarily contractions that eliminate redundancies.\footnote{Also, contractions do not yield disjoint state spaces. To obtain disjoint state spaces, we would have to consider additionally models isomorphic to contractions but with disjoint state spaces.} We are unsure of whether it is necessarily a natural property when we interpret FH models as subjective views of agents. An agent may not realize or may not be bothered by redundancies and use an FH model with redundancies to analyze her situation. That is, differences in awareness and redundancies are orthogonal to each other and reduction in awareness does not necessitate elimination of redundancies.

\section{Transformations}

\subsection{From FH Models to Complemented HMS Models}\label{sec: u-transf}

We can use the tools of the prior sections to define a transformation of a FH model into a complemented HMS model. The transformation works as follows: For any FH model $\mathsf{K}_{\mathsf{At}}$ for $\mathsf{At}$, consider the category of FH models $\langle (\mathsf{K}_{\Phi})_{\Phi \subseteq \mathsf{At}}, (f^{\Phi}_{\Psi})_{\Psi \subseteq \Phi \subseteq \mathsf{At}} \rangle$. This category is transformed into an implicit knowledge-based HMS model. We then derive the explicit possibility correspondences and add them to the implicit knowledge-based HMS model, obtaining a complemented implicit knowledge-based HMS model as in Section~\ref{implicit-to-explicit}. In the next step, we erase the awareness functions and get a complemented HMS model. The core step is to transform a category of FH models into an implicit knowledge-based HMS model. This is defined next.  

\begin{defin}[$T$-Transform]\label{T_transform} For any category of FH models $\mathcal{C}(\mathsf{K}_{\mathsf{At}}) = \langle (\mathsf{K}_{\Phi})_{\Phi \subseteq \mathsf{At}}, \\ (f^{\Phi}_{\Psi})_{\Psi \subseteq \Phi \subseteq \mathsf{At}} \rangle$, the \emph{$T$-transform} of $\mathcal{C}(\mathsf{K}_{\mathsf{At}})$ is the implicit knowledge-based HMS model $T(\mathcal{C}(\mathsf{K}_{\mathsf{At}})) = \langle I, \{S_{\Phi}\}_{\Phi \subseteq \mathsf{At}}, (r^{\Phi}_{\Psi})_{\Psi \subseteq \Phi \subseteq \mathsf{At}}, (\Lambda^*_i)_{i \in I}, (\alpha_i)_{i \in I}, v \rangle$ defined as follows: 
\begin{itemize}
\item $S_{\Phi} = W_{\Phi}$ for all $\Phi \subseteq \mathsf{At}$, where $W_\Phi$ is the state space of the FH model $\mathsf{K}_{\Phi}$ of the category $\mathcal{C}(\mathsf{K}_{\mathsf{At}})$. Denote $\Omega = \bigcup_{\Phi \subseteq \mathsf{A}} S_{\Phi}$.

\item $r^{\Phi}_{\Psi} = f^{\Phi}_{\Psi}$ for any $\Phi, \Psi \subseteq \mathsf{At}$ with $\Psi \subseteq \Phi$, where $f^{\Phi}_{\Psi}$ is the surjective bounded morphism of the category $\mathcal{C}(\mathsf{K}_{\mathsf{At}})$. 

\item $\Lambda^*_i: \Omega \longrightarrow 2^\Omega$ such that $\Phi \subseteq \mathsf{At}$ and $w \in S_{\Phi}$, $w' \in \Lambda^*_i(w)$ if and only if $(w, w') \in R_{\Phi,i}$, for any $i \in I$.

\item $\alpha_i: \Omega \longrightarrow \{S_{\Phi}\}_{\Phi \subseteq \mathsf{At}}$ such that for all $\Psi \subseteq \mathsf{At}$ and $w \in S_{\Psi}$, $\alpha_i(w) = S_{\Upsilon}$ if and only if $\mathsf{At}(\mathcal{A}_{\Psi, i}(w)) = \Upsilon$, for any $i \in I$.
 
\item  $v(p) = \bigcup_{\Phi \subseteq \mathsf{At}} V_{\Phi}(p)$, for any $p \in \mathsf{At}$.
\end{itemize}
\end{defin}

The $T$-transform indeed transforms any category of $FH$ models into an implicit knowledge-based HMS model. 

\begin{prop}\label{prop: T-transform}  For any category of FH models $\mathcal{C}(\mathsf{K}_{\mathsf{At}})$, the $T$-transform $T(\mathcal{C}(\mathsf{K}_{\mathsf{At}}))$ is an implicit knowledge-based HMS model.
\end{prop}

\noindent \textsc{Proof.}  Fix a category of $FH$ models, $\mathcal{C}(\mathsf{K}_{\mathsf{At}}) = \langle (\mathsf{K}_{\Phi})_{\Phi \subseteq \mathsf{At}}, (f^{\Phi}_{\Psi})_{\Psi \subseteq \Phi \subseteq \mathsf{At}} \rangle$ with $\mathsf{K}_{\Phi} = \langle I, W_{\Phi}, (R_{\Phi, i})_{i \in I}, (\mathcal{A}_{\Phi,i})_{i \in I}, V_{\Phi}\rangle$ for $\Phi \subseteq \mathsf{At}$. We show that we can obtain from this category each component of an implicit knowledge-based HMS model. 

\emph{Collection of state spaces:} $\{S_{\Phi}\}_{\Phi \subseteq \mathsf{At}}$ forms a complete lattice by set inclusion on $\mathsf{At}$. 

\emph{Projections:} Since $r^{\Phi}_{\Psi} = f^{\Phi}_{\Psi}$ for any $\Phi, \Psi \subseteq \mathsf{At}$ with $\Psi \subseteq \Phi$, the fact that projections commute follows immediately from $(f^{\Phi}_{\Psi})_{\Phi, \Psi \subseteq \mathsf{At}}$ being morphisms of the category $\mathcal{C}(\mathsf{K}_{\mathsf{At}})$. It also implies that $r^{\Phi}_{\Phi}$ is the identity on $S_{\Phi}$. The fact that $r^{\Phi}_{\Psi}$ is a surjection follows directly from $f^{\Phi}_{\Psi}$ being a surjective bounded morphism. 

\emph{Implicit possibility correspondences:} For every $\Phi \subseteq \mathsf{At}$, the partitional properties of $\Lambda^*_i$ restricted to $S_{\Phi}$ follow directly from the partial properties of $R_{\Phi, i}$. We need to show Projections Preserve Implicit Knowledge, that is, we need to show that for any $\Phi \subseteq \mathsf{At}$, if $\omega \in S_{\Phi}$, then $\Lambda^*_i(\omega)_{\Psi} = \Lambda^*(\omega_{\Psi})$. 

``$\subseteq$'': If $\omega' \in \Lambda^*_i(\omega)_{\Psi}$, then $(f^{\Phi}_{\Psi})^{-1}(\omega') \subseteq \Lambda^*_i(\omega)$ since $f^{\Phi}_{\Psi}$ is a surjection. For all $\omega'' \in (f^{\Phi}_{\Psi})^{-1}(\omega')$, $(\omega, \omega'') \in R_{\Phi, i}$ by $T$-transform. Since $f^{\Phi}_{\Psi}$ is a homomorphism, $(f^{\Phi}_{\Psi}(\omega), f^{\Phi}_{\Psi}(\omega'')) \in R_{\Psi, i}$. By the $T$-transform, $f^{\Phi}_{\Psi}(\omega'') \in \Lambda^*_i(\omega_{\Psi})$. Thus, $\omega' \in \Lambda^*_i(\omega_{\Psi})$.

``$\supseteq$'': If $\omega' \in \Lambda^*_i(\omega_{\Psi})$, then by the $T$-transform $(\omega_{\Psi}, \omega') \in R_{\Psi, i}$. By Back, there exists $t \in (f^{\Phi}_{\Psi})^{-1}(\omega')$ with $(\omega, t) \in R_{\Phi, i}$. By the $T$-transform, $t \in \Lambda^*_i(\omega)$. Hence, $\omega' \in \Lambda^*_i(\omega)_{\Psi}$. 

\emph{Awareness function:} We verify the properties one-by-one. 

0.: For all $w \in W_{\Phi}$, $\mathcal{A}_{\Phi, i}(w) \subseteq \mathcal{L}_{\Phi}$. Thus, $\mathsf{At}(\mathcal{A}_{\Phi, i}(w)) \subseteq \Phi$. Set $\Psi = \mathsf{At}(\mathcal{A}_{\Phi, i}(w))$. By $T$-transform, $\alpha_i(w) = S_{\Psi}$. 

I.: If $\omega' \in \Lambda^*_i(\omega)$, then by the $T$-transform $(\omega, \omega') \in R_{\Phi, i}$ for $\Phi \subseteq \mathsf{At}$ with $\omega' \in S_{\Phi}$. Since in a propositionally determined FH model, agents know what they are aware of, $\mathcal{A}_{\Phi, i}(\omega) = \mathcal{A}_{\Phi, i}(\omega')$. By the $T$-transform, $\alpha_i(\omega) = \alpha_i(\omega')$. 

II.: If $\omega \in S_{\Phi}$ and $S_{\Psi} \preceq \alpha_i(\omega)$, then by the $T$-transform $\mathsf{At}(\mathcal{A}_{\Phi, i}(\omega)) \supseteq \Psi$. By Awareness consistency, $\mathcal{A}_{\Psi, i}(f^{\Phi}_{\Psi}(\omega)) \cap \mathcal{L}_{\Psi}$. Thus, $\mathsf{At}(\mathcal{A}_{\Psi, i}(\omega_{\Psi})) = \Psi$. By the $T$-transform, $\alpha_i(\omega_{\Psi}) = S_{\Psi}$. 

III.: If $\omega \in S_{\Phi}$ and $\alpha_i(\omega) \preceq S_{\Psi} \preceq S_{\Phi}$, then by the $T$-transform $\mathsf{At}(\mathcal{A}_{\Phi, i}(\omega)) \subseteq \Psi \subseteq \Phi$. Denote $\Upsilon = \mathsf{At}(\mathcal{A}_{\Phi, i}(\omega))$. By Awareness consistency, $\mathcal{A}_{\Psi, i}(f^{\Phi}_{\Psi}(\omega)) = \mathcal{A}_{\Phi, i}(\omega) \cap \mathcal{L}_{\Psi} = \mathcal{L}_{\Upsilon}$. Thus, $\mathsf{At}(\mathcal{A}_{\Psi, i}(f^{\Phi}_{\Psi}(\omega))) = \Upsilon$. By the $T$-transform, $\alpha_i(\omega_{\Psi}) = \alpha_i(\omega)$. 

IV.: If $\omega \in S_{\Phi}$ and $\Psi \subseteq \Phi$, then by Awareness consistency $\mathcal{A}_{\Phi, i}(\omega) \cap \mathcal{L}_{\Psi} = \mathcal{A}_{\Psi, i}(f^{\Phi}_{\Psi}(\omega))$. Thus, $\mathsf{At}(\mathcal{A}_{\Phi, i}(\omega)) \supseteq \mathsf{At}(\mathcal{A}_{\Psi, i}(f^{\Phi}_{\Psi}(\omega)))$. Applying the $T$-transform, yields $\alpha_i(\omega) \succeq \alpha_i(\omega_{\Psi})$. 

\emph{Valuation:} We have to show that $v(p)$ is an event for every $p \in \mathsf{At}$. Take $S_{\{p\}}$ to be the base-space and $V_{\{p\}}(p)$ to be the base. Then it follows immediately from the definition of $T$-transform that $v(p)$ is an event. 


This completes the proof of the proposition.\hfill $\Box$\\

We have all ingredients to define the transformation of FH models into complemented HMS models. 

\begin{defin}[HMS-Transform]\label{HMS_transform} For any FH model $\mathsf{K}_{\mathsf{At}}$, the \emph{HMS-transform} $HMS(\mathsf{K}_{\mathsf{At}}) = \langle I, \{S_{\Phi}\}_{\Phi \subseteq \mathsf{At}}, (r^{\Phi}_{\Psi})_{\Psi \subseteq \Phi \subseteq \mathsf{At}}, (\Lambda^*_i)_{i \in I}, (\Pi^*_i)_{i \in I}, v \rangle$ is defined by the following steps:
\begin{enumerate}

\item Form the category of FH models $\mathcal{C}(\mathsf{K}_{\mathsf{At}})$ (Definition~\ref{category}).

\item Apply the $T$-transform to $\mathcal{C}(\mathsf{K}_{\mathsf{At}})$ to obtain the implicit knowledge-based HMS model $T(\mathcal{C}(\mathsf{K}_{\mathsf{At}})) = \langle I, \{S_{\Phi}\}_{\Phi \subseteq \mathsf{At}}, (r^{\Phi}_{\Psi})_{\Psi \subseteq \Phi \subseteq \mathsf{At}}, (\Lambda^*_i)_{i \in I}, (\alpha_i)_{i \in I}, v \rangle$ (Definition~\ref{T_transform}).

\item Form the complemented implicit knowledge-based HMS model $\overline{T(\mathcal{C}(\mathsf{K}_{\mathsf{At}}))} = \langle I, \{S_{\Phi}\}_{\Phi \subseteq \mathsf{At}}, \\ (r^{\Phi}_{\Psi})_{\Psi \subseteq \Phi \subseteq \mathsf{At}}, (\Lambda^*_i)_{i \in I}, (\Pi^*_i)_{i \in I}, (\alpha_i)_{i \in I}, v \rangle$ by deriving the explicit possibility correspondences $(\Pi^*_i)_{i \in I}$ (Definition~\ref{derived_explicit_possibility_correspondence}). 

\item Erase the awareness functions $(\alpha_i)_{i \in I}$ from the complemented implicit knowledge-based HMS model $\overline{T(\mathcal{C}(\mathsf{K}_{\mathsf{At}}))}$ to obtain the complemented HMS model $\langle I, \{S_{\Phi}\}_{\Phi \subseteq \mathsf{At}}, \\ (r^{\Phi}_{\Psi})_{\Psi \subseteq \Phi \subseteq \mathsf{At}}, (\Lambda^*_i)_{i \in I}, (\Pi^*_i)_{i \in I}, v \rangle$. 

\end{enumerate}
\end{defin}

\begin{cor} \label{cor: transform2} For any FH model $\mathsf{K}_{\mathsf{At}}$, its HMS-transform $HMS(\mathsf{K}_{\mathsf{At}})$ is a complemented HMS model.
\end{cor}

\noindent \textsc{Proof.}  The proof is a corollary from the preceding results. For any  $\mathsf{K}_{\mathsf{At}}$, consider the category $\mathcal{C}(\mathsf{K}_{\mathsf{At}})$ (Definition~\ref{category}). By Proposition~\ref{prop: T-transform}, the $T$-transform of  $\mathcal{C}(\mathsf{K}_{\mathsf{At}})$ yields the implicit knowledge-based model $T(\mathcal{C}(\mathsf{K}_{\mathsf{At}})) = \langle I, \{S_{\Phi}\}_{\Phi \subseteq \mathsf{At}}, (r^{\Phi}_{\Psi})_{\Psi \subseteq \Phi \subseteq \mathsf{At}}, (\Lambda^*_i)_{i \in I}, (\alpha_i)_{i \in I}, v \rangle$. Use Definition~\ref{derived_explicit_possibility_correspondence} to derive $\Pi_i^*$ for every $i \in I$. By Lemmata~\ref{immediate} to~\ref{measurability_return}, $\Pi_i^*$ is indeed an explicit possibility correspondence for every $i \in I$ and $\overline{T(\mathcal{C}(\mathsf{K}_{\mathsf{At}}))} = \langle I, \{S_{\Phi}\}_{\Phi \subseteq \mathsf{At}}, (r^{\Phi}_{\Psi})_{\Psi \subseteq \Phi \subseteq \mathsf{At}}, (\Lambda^*_i)_{i \in I}, \\ (\Pi^*_i)_{i \in I}, (\alpha_i)_{i \in I}, v \rangle$ is a complemented implicit knowledge-based HMS model. Erasing the awareness functions $(\alpha_i)_{i \in I}$ yields an complemented HMS model $\langle I, \{S_{\Phi}\}_{\Phi \subseteq \mathsf{At}}, (r^{\Phi}_{\Psi})_{\Psi \subseteq \Phi \subseteq \mathsf{At}}, \\ (\Lambda^*_i)_{i \in I}, (\Pi^*_i)_{i \in I}, v \rangle$ by Corollary~\ref{cHMS_derived}. This is the HMS-transform $HMS(\mathsf{K}_{\mathsf{At}})$. \hfill $\Box$

\subsection{From Complemented HMS Models to FH Models}\label{sec: fh-transf}

To transform a complemented HMS model into a FH model we simply need to consider the upmost space of the lattice of spaces of the HMS model, copy  the domain, define accessibility relations from implicit possibility correspondences as well as the valuation function, and, for every state $\w \in S_{At}$, construct the awareness set at $\w$ by collecting all the formulas that contain the atoms defined in the space where $\Pi_i(\w)$ lies. 
	
\begin{defin}[FH-Transform] For any complemented HMS model $\overline{\mathsf{M}} = \langle I, \{S_{\Phi}\}_{\Phi \subseteq \mathsf{At}}, \\ (r^{\Phi}_{\Psi})_{\Psi \subseteq \Phi \subseteq \mathsf{At}}, (\Lambda_i)_{i \in I}, (\Pi_i)_{i \in I}, v \rangle$, the  \emph{$FH$-transform} $FH(\overline{\mathsf{M}}) = \langle I, W_{\mathsf{At}}, (R_{\mathsf{At}, i})_{i \in I}, (\mathcal{A}_{\mathsf{At},i})_{i \in I}, V_{\mathsf{At}}\rangle$ is defined by 
\begin{itemize}
\item $W_{\mathsf{At}} = S_{\mathsf{At}},$
\item $R_{\mathsf{At}, i} \subseteq W_{\mathsf{At}} \times W_{\mathsf{At}}$ is such that $(\w, \w') \in R_{\mathsf{At}, i}$ if and only if $\w' \in \Lambda_i(\w)$, for all $i \in I$, 
\item $\mathcal{A}_{\mathsf{At}, i}: W_{\mathsf{At}} \longrightarrow 2^{\mathcal{L}_{\mathsf{At}}}$ is such that $\mathcal{A}_{\mathsf{At}, i} (\w) = \mathcal{L}_{\Phi}$ for $\Phi \subseteq \mathsf{At}$ with $\Pi_i(\w) \subseteq S_{\Phi}$, for all $i \in I$,
\item $V_{\mathsf{At}}: \mathsf{At} \longrightarrow 2^{W_{\mathsf{At}}}$ is such that $V_{\mathsf{At}}(p) = v(p) \cap S_{\mathsf{At}}$, for every $p \in \mathsf{At}$.
\end{itemize}
\end{defin}

The FH-transform indeed transforms any complemented HMS model into a FH model. 

\begin{prop}\label{prop: transform1} For every complemented HMS model $\overline{\mathsf{M}}$, the FH-transform $FH(\overline{\mathsf{M}})$ is a FH model for $\mathsf{At}$.
\end{prop}

\noindent \textsc{Proof.} The partitional properties of $R_{\mathsf{At}, i}$ follow in a straightforward way from the partitional properties of $\Lambda_i$ on $S_{\mathsf{At}}$, for every $i \in I$. 

The fact that Awareness is Generated by Primitive Propositions follows directly from the clause of the FH-transform pertaining to the awareness correspondence $\mathcal{A}_{\mathsf{At}, i}$. What is left to show is that Agents Know What They Are Aware of. The complemented HMS model satisfies Explicit Measurability. This means in particular that for all $\omega \in S_{\mathsf{At}}$, $\omega' \in \Lambda_i(\omega)$, we have $\Pi_i(\omega') = \Pi_i(\omega)$. By the clause of the FH-transform pertaining to $R_{\mathsf{At}, i}$, it follows immediately that $\omega' \in \Lambda_i(\omega)$ implies $(\omega, \omega') \in R_{\mathsf{At}, i}$. By the clause of the FH-transform pertaining to the awareness correspondence $\mathcal{A}_{\mathsf{At}, i}$, we have now that $\Pi_i(\omega') = \Pi_i(\omega)$ implies $\mathcal{A}_{\mathsf{At}, i}(\omega') = \mathcal{A}_{\mathsf{At}, i}(\omega)$.\hfill$\Box$

\subsection{Equivalence of Complemented HMS and FH Models\label{subsec: equivalence LKA}}

Before we can prove an equivalence of HMS and FH models, we need to introduce the semantics of complemented HMS models. 

\begin{defin}[Semantics of HMS Models]\label{def: satisfiability of L} Let $\overline{\mathsf{M}} = \langle I, \{S_{\Phi}\}_{\Phi \subseteq \mathsf{At}}, (r^{\Phi}_{\Psi})_{\Psi \subseteq \Phi \subseteq \mathsf{At}}, (\Lambda_i)_{i \in I}, \\ (\Pi_i)_{i \in I}, v \rangle$ be a complemented HMS model and let $\w \in \Omega$. Satisfaction of $\mathcal{L}_{\mathsf{At}}$ formulas in $\overline{\mathsf{M}}$ is given by \ $\overline{\mathsf{M}},\w \vDash \top$ for all $\w \in \Omega$ and
\begin{center}
	\begin{tabular}{lllcclll}
	$\overline{\mathsf{M}},\w \vDash p$ & $ \text{ iff }$ & $\w \in v(p)$; & & & $\overline{\mathsf{M}}, \w \vDash a_i \varphi$ & $\text{ iff } $& $S_{\Pi_i(\w)} \succeq S([\varphi])$; \tabularnewline
	$\overline{\mathsf{M}}, \w \vDash \neg \varphi$ & $\text{ iff }$ & $\w \in \neg [\varphi]$; & & & $\overline{\mathsf{M}},\w \vDash \ell_i\varphi$ & $ \text{ iff } $& $\Lambda_i(\w)\subseteq [\varphi]$; \tabularnewline
	$\overline{\mathsf{M}}, \w \vDash \varphi \wedge \psi $& $\text{ iff }$ & $ \w \in [\varphi] \cap [\psi]$; & & & $\overline{\mathsf{M}},\w \vDash k_i \varphi$ & $ \text{ iff }$ & $\Pi_i(\w) \subseteq [\varphi]$;\tabularnewline
\end{tabular}
\end{center}
\noindent where $[\varphi] := \{\w' \in \Omega : \overline{\mathsf{M}}, \w' \vDash \varphi\}$ for all $\varphi \in \mathcal{L}_{\mathsf{At}}$.
\end{defin}

A couple of comments are in order: First, in HMS models, formulas may have undefined truth value since a formula may not be defined in every state. The same happens in FH models of a category of FH models. For instance, the truth value of $p$ is not defined for all FH models $\mathsf{K}_{\Phi}$ with $\Phi \not\ni p$. We will return to this issue later in Section~\ref{soundness_completeness} when we prove soundness and completeness. Second, recall that for all $p\in \mathsf{At}, v(p)$ is an event, so $[p]$ is an event in $\Sigma$. Negation and intersection of events are events. By Lemmata~\ref{alles_events} and~\ref{Levent}, explicit knowledge, awareness, and implicit knowledge of events are also events, respectively. Thus, for every $\varphi \in \mathcal{L}_{\mathsf{At}}$, $[\varphi]$ is an event.

Proposition \ref{prop: semantic equivalence of K - L and A} shows that in complemented HMS models, $K_i(E) = L_i(E) \cap A_i(E)$, for any event $E \in {\Sigma}$, so the semantics of $\mathcal{L}_{\mathsf{At}}$ provided above immediately implies that:

\begin{prop}\label{prop: k iff L&A} For any complemented HMS model $\overline{\mathsf{M}}$, $\w \in \Omega$, $\varphi \in \mathcal{L}_{\mathsf{At}}$, and $\Psi \subseteq \mathsf{At}$ with $\mathsf{At}(\varphi) \subseteq \Psi$,
$$\overline{\mathsf{M}},\w_{\Psi} \vDash k_i \varphi \leftrightarrow (\ell_i \varphi \wedge a_i \varphi).$$
\end{prop}

An FH model and its HMS-transform satisfy the same formulas in the language $\mathcal{L}_{\mathsf{At}}$ with implicit knowledge, explicit knowledge, and awareness as long as these formulas are defined at the corresponding states of the HMS-transform. 

\begin{prop}\label{prop: equivalence2} For any FH model $\mathsf{K}_{\mathsf{At}}$ and its HMS-transform $HMS(\mathsf{K}_{\mathsf{At}})$, 
for all $w \in W_{\mathsf{At}}$, $\varphi \in \mathcal{L}_{\mathsf{At}}$, and $\Phi \subseteq \mathsf{At}$ with $\mathsf{At}(\varphi) \subseteq \Phi$, $$\mathsf{K}_{\mathsf{At}}, w \Vdash \varphi \mbox{ if and only if } HMS(\mathsf{K}_{\mathsf{At}}), w_\Phi \vDash \varphi.$$
\end{prop}

\noindent \textsc{Proof. } The proof proceeds by induction on the complexity of the formulas in $\mathcal{L}_{\mathsf{At}}$. The case $\top$ is trivial. 

\noindent Base Case: Consider now any $p \in \mathsf{At}$. We have that $\mathsf{K}_{\mathsf{At}}, w \Vdash p$ if and only if $w \in V_{\mathsf{At}}(p)$ if and only if (by the Atomic harmony clause of bounded morphisms) $f^{\mathsf{At}}_{\Phi}(w) \in V_{\Phi}(p)$ for all $\Phi \ni p$ if and only if $w_{\Phi} \in v(p)$ for all $\Phi \ni p$ (by the $T$-transform) if and only if $HMS(\mathsf{K}_{\mathsf{At}}), w_\Phi \vDash p$ for all $\Phi \ni p$. 

\noindent Inductive Step: The cases $\neg \varphi$ and $\varphi \wedge \psi$ are now straightforward. The requirement is now for all $\Phi \supseteq \mathsf{At}(\varphi)$ and $\Phi \supseteq \mathsf{At}(\varphi \wedge \psi)$, respectively. 

Now consider the case $\ell_i \varphi$: We have $\mathsf{K}_{\mathsf{At}}, w \Vdash \ell_i \varphi$ if and only if $\mathsf{K}_{\mathsf{At}}, t \Vdash \varphi$ for all $t \in W_{\mathsf{At}}$ with $(w, t) \in R_{\mathsf{At}, i}$ if and only if (by the properties of bounded morphisms) $\mathsf{K}_{\mathsf{\Phi}}, f^{\mathsf{At}}_{\Phi}(t) \Vdash \varphi$ for all $f^{\mathsf{At}}_{\Phi}(t) \in W_{\mathsf{\Phi}}$ with $(f^{\mathsf{At}}_{\Phi}(w), f^{\mathsf{At}}_{\Phi}(t)) \in R_{\mathsf{\Phi}, i}$ for all $\Phi \subseteq \mathsf{At}$ with $\mathsf{At}(\varphi) \subseteq \Phi$ if and only if (by the induction hypothesis and the $T$-transform) $HMS(\mathsf{K}_{\mathsf{\Phi}}), t_{\Phi} \vDash \varphi$ for all $t_{\Phi} \in W_{\mathsf{\Phi}}$ with $t_{\Phi} \in \Lambda_i(w_{\Phi})$ for all $\Phi \subseteq \mathsf{At}$ with $\mathsf{At}(\varphi) \subseteq \Phi$ if and only if $HMS(\mathsf{K}_{\mathsf{\Phi}}), t_{\Phi} \vDash \ell_i \varphi$. 

Next, consider the case $\alpha_i \varphi$: We have $\mathsf{K}_{\mathsf{At}}, w \Vdash a_i \varphi$ if and only if $\varphi \in \mathcal{A}_{\mathsf{At}, i}(w)$ if and only if (by Awareness consistency of bounded morphisms) $\varphi \in \mathcal{A}_{\mathsf{At}, i}(w) \cap \mathcal{L}_{\Phi} = \mathcal{A}_{\mathsf{\Phi}, i}(f^{\mathsf{At}}_{\Phi} (w))$ for all $\Phi \subseteq \mathsf{At}$ with $\mathsf{At}(\varphi) \subseteq \Phi$ if and only if (by the $T$-transform) $\alpha_i(w_{\Phi}) \succeq S([\varphi])$ for all $\Phi \ni \varphi$ if and only if $HMS(\mathsf{K}_{\mathsf{At}}), w_{\Phi} \vDash a_i \varphi$ for all $\Phi \subseteq \mathsf{At}$ with $\mathsf{At}(\varphi) \subseteq \Phi$. 

Finally, the case $k_i \varphi$ follows now by the induction hypothesis, $k_i \varphi \leftrightarrow (\ell_i \varphi \wedge a_i \varphi)$, and Proposition~\ref{prop: k iff L&A}. \hfill $\Box$\\

Conversely, we now show that any complemented HMS model and its FH-transform satisfy the same formulas from the language $\mathcal{L}_{\mathsf{At}}$ with implicit knowledge, explicit knowledge, and awareness.  

\begin{prop}\label{prop: equivalence1} For any complemented HMS model $\overline{\mathsf{M}}$ and its FH-transform $FH(\overline{\mathsf{M}})$, for all $\varphi \in\mathcal{L}_{\mathsf{At}}$ and all $\w \in S_{At}$,
$\overline{\mathsf{M}}, \w \vDash \varphi$ if and only if $FH(\overline{\mathsf{M}}),\w \Vdash \varphi.$
\end{prop}

\noindent \textsc{Proof.}  The proof proceeds by straightforward induction on the complexity of formulas in $\mathcal{L}_{\mathsf{At}}$ starting with the trivial case $\top$, the base case $p \in \mathsf{At}$, followed by the inductive step for $\neg \varphi$, $\varphi \wedge \psi$, $\ell_i \varphi$, and $a_i \varphi$. Note that the case $k_i \varphi$ follows now by Proposition~\ref{prop: k iff L&A} from the preceding steps.\hfill $\Box$


\subsection{From FH to Implicit Knowledge-based HMS Models}\label{sec: u*-transf}

We now focus on the relationship between implicit knowledge-based HMS and FH models. This relationship is even simpler than for completed HMS and FH models since implicit knowledge-based HMS models are arguably already closer to FH models than complemented HMS models. This is due to taking implicit knowledge and the awareness functions as primitives. 

We define a version of HMS transformation that is ``truncated'' after the $T$-transformation. It just keeps the first two steps of the HMS transformation. 

\begin{defin}[Truncated HMS-Transform] For any FH model $\mathsf{K}_{\mathsf{At}}$, the \emph{truncated HMS-transform} $HMS^*(\mathsf{K}_{\mathsf{At}}) = \langle I, \{S_{\Phi}\}_{\Phi \subseteq \mathsf{At}}, (r^{\Phi}_{\Psi})_{\Psi \subseteq \Phi \subseteq \mathsf{At}}, (\Lambda^*_i)_{i \in I}, (\alpha^*_i)_{i \in I}, v \rangle$ is defined by the first two steps of the HMS-transform (cf. Definition~\ref{HMS_transform}). 
\end{defin}

From Proposition~\ref{prop: T-transform} follows now immediately: 

\begin{cor}\label{prop: trunctated HMS-transform} For any FH model $\mathsf{K}_{\mathsf{At}}$, the truncated HMS-transform $HMS^*(\mathsf{K}_{\mathsf{At}})$ is an implicit knowledge-based HMS model.
\end{cor}

\subsection{From Implicit Knowledge-based HMS to FH Models}\label{sec: fh*-transf}

\begin{defin}[FH$^*$-Transform] For any implicit knowledge-based HMS model $\overline{\mathsf{M}}^* = \langle I, \{S_{\Phi}\}_{\Phi \subseteq \mathsf{At}}, (r^{\Phi}_{\Psi})_{\Psi \subseteq \Phi \subseteq \mathsf{At}}, (\Lambda_i)_{i \in I}, (\alpha_i)_{i \in I}, v \rangle$, the  \emph{FH$^*$-transform} $FH^*(\overline{\mathsf{M}}^*) = \langle I, W_{\mathsf{At}}, \\ (R_{\mathsf{At}, i})_{i \in I}, (\mathcal{A}_{\mathsf{At},i})_{i \in I}, V_{\mathsf{At}}\rangle$ is defined like the $FH$-transform except that the clause for the awareness correspondence is replaced by, for any $i \in I$, 
\begin{itemize} 
\item $\mathcal{A}_{\mathsf{At}, i}: W_{\mathsf{At}} \longrightarrow 2^{\mathcal{L}_{\mathsf{At}}}$ is such that $\mathcal{A}_{\mathsf{At}, i} (\w) = \mathcal{L}_{\Phi}$ for $\Phi \subseteq \mathsf{At}$ with $\alpha_i(\w) = S_{\Phi}$. 
\end{itemize}
\end{defin}

The FH$^*$-transform indeed transforms any implicit knowledge-based HMS model into a FH model. 

\begin{prop}\label{prop: transform1*} For every implicit knowledge-based HMS model $\overline{\mathsf{M}}^*$, the FH$^*$-transform $FH^*(\overline{\mathsf{M}}^*)$ is a FH model for $\mathsf{At}$.
\end{prop}

\noindent \textsc{Proof.} The proof is analogous to the proof of Proposition~\ref{prop: transform1} except we need to modify the clause relating to the awareness correspondence. 

The fact that Awareness is Generated by Primitive Propositions follows directly from the clause of the FH$^*$-transform pertaining the awareness correspondence $\mathcal{A}_{\mathsf{At}, i}$. What is left to show is that Agents Know What They Are Aware of. The implicit knowledge-based HMS model satisfies Awareness Measurability. This means in particular that for all $\omega \in S_{\mathsf{At}}$, $\omega' \in \Lambda_i(\omega)$, we have $\alpha_i(\omega') = \alpha_i(\omega)$. By the clause of the FH$^*$-transform pertaining to $R_{\mathsf{At}, i}$, it follows immediately that $\omega' \in \Lambda_i(\omega)$ implies $(\omega, \omega') \in R_{\mathsf{At}, i}$. By the clause of the FH$^*$-transform pertaining the awareness correspondence $\mathcal{A}_{\mathsf{At}, i}$, we have now that $\alpha_i(\omega') = \alpha_i(\omega)$ implies $\mathcal{A}_{\mathsf{At}, i}(\omega') = \mathcal{A}_{\mathsf{At}, i}(\omega)$.\hfill$\Box$

\subsection{Equivalence of Implicit Knowledge-based HMS and FH Models} 

To prove the equivalence, we need to introduce the semantics of $\mathcal{L}_{At}$ over implicit knowledge-based HMS models. The semantics is identical to the semantics over complemented HMS models, except for the awareness clause.\footnote{We slightly abuse notation and use the same symbol for the satisfaction relation as we did for complemented HMS models.}

\begin{defin}\label{def: satisfiability of L2} Satisfaction of $\mathcal{L}_{\mathsf{At}}$ formulas in an implicit knowledge-based HMS model $\mathsf{M}^*$ is given like for complemented HMS models except that we have $\mathsf{M}^*, \w \vDash a_i \varphi$ if and only if $\alpha_i(\w) \succeq S([\varphi])$.
\end{defin} 

From this semantics and the syntactic definition $k_i \varphi := \ell_i \varphi \wedge a_i \varphi$, it follows that $\mathsf{M}^*,\w \vDash k_i\varphi$ if and only if $\mathsf{M}^*,\w \vDash \ell_i\varphi$ and $\mathsf{M}^*,\w \vDash a_i\varphi$. The comments about the semantics of complemented HMS models from Section~\ref{subsec: equivalence LKA} (right below the semantics definition) as well as Proposition~\ref{prop: k iff L&A}, also hold for implicit knowledge-based HMS models. 

An FH model and its truncated HMS-transform satisfy the same formulas in the language $\mathcal{L}_{\mathsf{At}}$ with implicit knowledge, explicit knowledge, and awareness with the provision that these formulas are defined at the corresponding states of the implicit knowledge-based HMS-transform. This follows directly from the proof of Proposition~\ref{prop: equivalence2}.\footnote{The proof of Proposition~\ref{prop: equivalence2} makes use of the HMS-transform, which in turn makes use of the $T$-transform. When arguments in that proof are stopped after making use of the $T$-transform, the proof of the corollary is completed.}  

\begin{cor}\label{prop: equivalence2*} For any FH model $\mathsf{K}_{\mathsf{At}}$ and its HMS-transform $HMS(\mathsf{K}_{\mathsf{At}})$, 
for all $w \in W_{\mathsf{At}}$, $\varphi \in \mathcal{L}_{\mathsf{At}}$, and $\Phi \subseteq \mathsf{At}$ with $\mathsf{At}(\varphi) \subseteq \Phi$, $$\mathsf{K}_{\mathsf{At}}, w \Vdash \varphi \mbox{ if and only if } HMS(\mathsf{K}_{\mathsf{At}}), w_\Phi \vDash \varphi.$$
\end{cor}

Any implicit knowledge-based HMS model and its FH$^*$-transform satisfy the same formulas from the language $\mathcal{L}_{\mathsf{At}}$ with implicit knowledge, explicit knowledge, and awareness. 

\begin{prop}\label{prop: equivalence1*} For any implicit knowledge-based HMS model $\mathsf{M}^*$ and its FH$^*$-transform $FH^*(\mathsf{M}^*)$, for all $\varphi \in\mathcal{L}_{\mathsf{At}}$ and all $\w \in S_{At}$,
$$\mathsf{M}^*, \w \vDash \varphi \mbox{ if and only if } FH^*(\mathsf{M}^*),\w \Vdash \varphi.$$
\end{prop}

\noindent \textsc{Proof.}  The proof proceeds by straightforward induction on the complexity of formulas in $\mathcal{L}_{\mathsf{At}}$ starting with the trivial case $\top$, the base case $p \in \mathsf{At}$, followed by the inductive step for $\neg \varphi$, $\varphi \wedge \psi$, $\ell_i \varphi$, $a_i \varphi$, and $k_i \varphi$.\hfill $\Box$

\section{Logic of Propositional Awareness\label{soundness_completeness}}

In this section, we explore the implications of the prior sections for axiomatizations of both the category of FH models and HMS models (both complemented and implicit knowledge-based). In particular, we show that the Logic of Propositional Awareness is sound and complete with respect to the class of complemented HMS models. This is the first axiomatization of HMS models that features also the notion of implicit knowledge. Previous axiomatizations of HMS models (Heifetz, Meier, and Schipper, 2008, Halpern and R\^{e}go, 2008) were confined to explicit knowledge and awareness only as the model did not contain a notion of implicit knowledge. We also show that the Logic of Propositional Awareness is sound and complete with respect to the class of implicit knowledge-based HMS models. This is the first axiomatization of implicit knowledge-based HMS models. Finally, it is also sound and complete with respect to the class of categories of FH models. 

\begin{defin} The logic LPA is the smallest set of $\mathcal{L}_{\mathsf{At}}$ formulas that contains the axioms in, and is closed under the inference rules of, Table \ref{tab:logic_LPA}.
\end{defin}

\begin{table}\caption{\label{tab:logic_LPA} Axioms and inference rules of the Logic of Propositional Awareness (LPA), for a propositionally determined notion of awareness and axioms for an $S5$ logic.}
	\begin{tabular}{|>{\raggedright}p{0.98\textwidth}|}
		\hline
		{\small{}All substitution instances of propositional logic, including
			the formula $\top$}{\small\par}
		
		{\small{}$(\ell_i\varphi\wedge(\ell_i\varphi\rightarrow \ell_i\psi))\rightarrow \ell_i\psi$\hfill{}(K,
			Distribution)}{\small\par}
		
		{\small{}$k_i\varphi\leftrightarrow(\ell_i\varphi\wedge a_i\varphi)$\hfill{}
			(Explicit Knowledge)}{\small\par}
		
		{\small{}$a_i(\varphi\wedge\psi)\leftrightarrow(a_i\varphi\wedge a_i\psi)$\hfill{}
			(A1, Awareness Distribution)}{\small\par}
		
		{\small{}$a_i\neg\varphi\leftrightarrow a_i\varphi$\hfill{}
			(A2, Symmetry)}{\small\par}
		
		{\small{}$a_ik_j\varphi\leftrightarrow a_i\varphi$\hfill{}
			(A3, Awareness of Explicit Knowledge)}{\small\par}
		
		{\small{}$a_ia_j\varphi\leftrightarrow a_i\varphi$\hfill{}
			(A4, Awareness Reflection)}{\small\par}
		
		{\small{}$a_i\ell_j\varphi\leftrightarrow a_i\varphi$\hfill{}
			(A5, Awareness of Implicit Knowledge)}{\small\par}
		
		{\small{}$a_i\varphi\rightarrow \ell_ia_i\varphi$\hfill{} (A11,
			Awareness Introspection)}{\small\par}
		
		{\small{}$\neg a_i\varphi\rightarrow \ell_i\neg a_i\varphi$\hfill{}
			(A12, Unawareness Introspection)}{\small\par}
		
		{\small{}From $\varphi$ and $\varphi\rightarrow\psi$, infer $\psi$
			\hfill{}(Modus Ponens)}{\small\par}
		
		{\small{}From $\varphi$ infer $\ell_i\varphi$ \hfill{}(K-Inference)}\tabularnewline
		
		{\small{}$\ell_i\varphi\rightarrow\varphi$\hfill{}(T, Truth)}{\small\par}
		
		{\small{}$\ell_i\varphi\rightarrow \ell_i\ell_i\varphi$\hfill{}(4,
			Positive Introspection)}{\small\par}
		
		{\small{}$\neg \ell_i\varphi\rightarrow \ell_i\neg \ell_i\varphi$\hfill{}(5,
			Negative Introspection)}\tabularnewline
		\hline
	\end{tabular}
\end{table}

In Table~\ref{tab:logic_LPA}, the Explicit Knowledge axiom captures the definition of explicit knowledge proposed by Fagin and Halpern (1988). Axioms 1-5 capture a propositionally generated notion of awareness, while axioms 11-12 characterize individuals know what they are aware of (the numbering of awareness axioms is taken from Halpern, 2001). Axioms K, T, 4, and 5 constitute the standard notion of S5 knowledge (see e.g., Rendsvig and Symons, 2021). Lastly, Table~\ref{tab:logic_LPA} contains inference rules and substitution that are standard in modal logic.
		
Recall that in a Kripke model or FH model, a formula is valid if it is true in every state. However, a formula $\varphi \in \mathcal{L}_{\mathsf{At}}$ is not even defined at states of the FH model $\mathsf{K}_{\Psi}$ with $\mathsf{At}(\varphi) \nsubseteq \Psi$. Similarly, as we remarked earlier when introducing the semantics for HMS models, a formula may not be defined in every state of a HMS model. We say that $\varphi$ is defined in state $\omega$ in the complemented HMS model $\overline{\mathsf{M}}$ if $\omega \in \bigcap_{p \in \mathsf{At}(\varphi)} (v(p) \cup \neg v(p))$ (and analogously for implicit knowledge-based HMS models). Similarly, we say that $\varphi$ is defined in the FH model $\mathsf{K}_{\Psi}$ if $\mathsf{At}(\varphi) \subseteq \Psi$. 

Now we say that a formula $\varphi$ is valid in the complemented HMS model $\overline{M}$ if $\overline{M}, \omega \vDash \varphi$ for all $\omega$ in which $\varphi$ is defined  (and analogously for the implicit knowledge-based HMS model). Similarly, we say that $\varphi$ is valid in the category of FH models $\mathcal{C}(\mathsf{K}_{\Psi})$ if $\mathsf{K}_{\Psi}, w \Vdash \varphi$ for all $w \in W_{\Psi}$ for all $\mathsf{K}_{\Psi}$ in $\mathcal{C}(\mathsf{K}_{\mathsf{At}})$ for which $\varphi$ is defined. A formula is valid in a class of complemented HMS models $\mathcal{M}$ if it is valid in every complemented HMS model of the class  (and analogously for the class of implicit knowledge-based HMS models). A formula is valid in a class of categories of FH models $\frak{C}$ if it is valid in every category of FH models of the class. 

A proof in an axiom system consists of a sequence of formulas, where each formula in the sequence is either an axiom in the axiom system or follows from the prior formula in the sequence by an application of an inference rule of the axiom system. A proof of a formula $\varphi$ is a proof where the last formula of the sequence is $\varphi$. A formula $\varphi$ is provable in an axiom system, if there is a proof of $\varphi$ in the axiom system. An axiom system is sound for the language $\mathcal{L}_{\mathsf{At}}$ with respect to a class of complemented HMS models $\mathcal{M}$ if every formula in $\mathcal{L}_{\mathsf{At}}$ that is provable in the axiom system is valid in every complemented HMS model of the class $\mathcal{M}$ (and analogously for the class of implicit knowledge-based HMS models). Similarly, an axiom system is sound for the language $\mathcal{L}_{\mathsf{At}}$ with respect to a class of categories of FH models $\frak{C}$ if every formula in $\mathcal{L}_{\mathsf{At}}$ that is provable in the axiom system is valid in every category of FH models of the class $\frak{C}$. An axiom system is complete for the language $\mathcal{L}_{\mathsf{At}}$ with respect to a class of complemented HMS models $\mathcal{M}$ if every formula in $\mathcal{L}_{\mathsf{At}}$ that is valid in $\mathcal{M}$ is provable in the axiom system (and analogously for the class of implicit knowledge-based HMS models). Similarly, an axiom system is complete for the language $\mathcal{L}_{\mathsf{At}}$ with respect to a class of categories of FH models $\frak{C}$ if every formula in $\mathcal{L}_{\mathsf{At}}$ that is valid in $\frak{C}$ is provable in the axiom system.

\begin{cor}\label{class of categories axiomatization} LPA is sound and complete with respect to
\begin{enumerate} 
\item the class of categories of FH models,
\item the class of complemented HMS models, 
\item the class of implicit knowledge-based HMS models.
\end{enumerate}   
\end{cor}

Fagin and Halpern (1988), Halpern (2001), and Halpern and R\^{e}go (2008) claim that LPA is sound and complete with respect to the class of FH models. The proof of (1.) now follows from invariance of modal satisfaction relative to sublanguages between FH models in each category of FH models, i.e., Proposition~\ref{epistemic_equivalence}. The proof of (2.) follows from Propositions~\ref{prop: equivalence2} and~\ref{prop: equivalence1}. The proof of (3.) follows from Corollary~\ref{prop: equivalence2*} and Proposition~\ref{prop: equivalence1*}.

\section{Discussion}

We enriched HMS models with a notion of implicit knowledge. We showed how implicit knowledge can be defined in HMS models so that it is consistent with explicit knowledge. We also introduced a variant of HMS models based on implicit knowledge and awareness and showed how to derive explicit knowledge. This demonstrates that explicit knowledge and implicit knowledge are in some sense interdefinable in HMS models. By introducing implicit knowledge into HMS models, we arguably made them ``closer'' to FH models. At the same time, we made FH models ``closer'' to HMS models by forming a category of FH models that differ by the language with bounded morphisms between them. This allowed us to consider subjective FH models, i.e., models that agents themselves could use at particular states to analyze their situation. 

The constructions also allowed us to consider the relation between FH models and HMS models, not just with respect to explicit knowledge and awareness, as in the prior literature, but also with respect to implicit knowledge. We showed an equivalence between FH and HMS models by transforming a model into the other and vice versa, and by showing that each model and its transform satisfy the same formulas. This equivalence is used to show that the Logic of Propositional Awareness is sound and complete with respect to the class of HMS models. Compared to the prior literature, this axiomatization is now for a language that also features implicit knowledge. 

\begin{figure}\caption{Relations between Approaches to Awareness\label{relations}}
\begin{center}
\includegraphics[scale = 0.15]{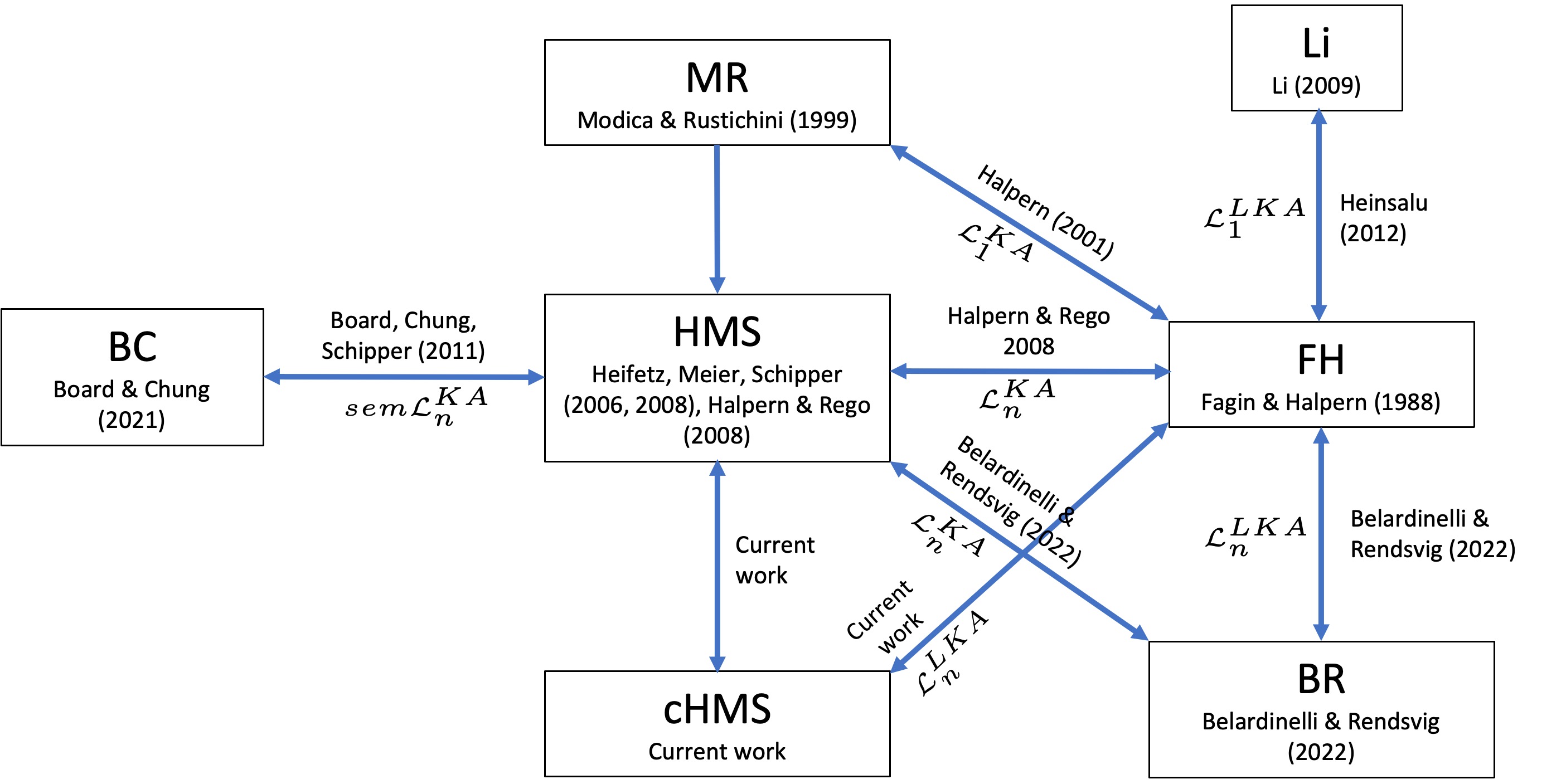} 
\end{center}
\end{figure}

The relations between various models of awareness in the literature are depicted in Figure~\ref{relations}. Beside FH models of Fagin and Halpern (1988) and HMS models of Heifetz, Meier, and Schipper (2006, 2008), we consider generalized standard models by Modica and Rustichini (1999), information structures with unawareness by Li (2009), object-based unawareness models by Board and Chung (2021), and Kripke lattices by Belardinelli and Rendsvig (2022). Equivalences hold for various languages also shown in the figure. We indicate the implicit, explicit, and awareness modality by superscripts $L$, $K$, and $A$, respectively. Some structures like Modica and Rustichini (1999) and Li (2009) feature just a single agent. We indicate this with the subscript ``$1$'' for single agent and ``$n$'' for multiple agents. For instance, $\mathcal{L}_n^{L, K, A}$ is the language featuring multiple agents, implicit knowledge, explicit knowledge, and awareness. The equivalence between Board and Chung (2021) is shown only at the level of semantics, i.e., at the level of events. The relation between Modica and Rustichini (1999) and Heifetz, Meier, and Schipper (2008) indicates that latter axiomatization can be seen as a multi-agent version of former. All shown relations pertain to rich structures featuring partitional knowledge and awareness generated by primitive propositions.

Recently, Schipper (2022) extended HMS models to awareness of unawareness by introducing quantified events. It would be straightforward to complement his structure with implicit knowledge as defined in the current work. Agents could then reason about the existence of their own implicit knowledge that they are not aware of. Such reasoning appears to be similar to the notion of speculative knowledge in van Ditmarsch et al. (2018). Awareness-of-unawareness structures with implicit knowledge would also allow for a better comparison to awareness structures with quantification of formulas for modeling reasoning about knowledge of unawareness (Halpern and R\^ego, 2009, 2012), object-based unawareness (Board and Chung, 2021), and quantified neighborhood structures with awareness (Sillari, 2008), all of which feature notions of implicit knowledge. 

As suggested in the introductory dialogue, we view our work as a first step towards a more comprehensive study of notions of implicit knowledge. In follow up work (Belardinelli and Schipper, 2024), we model tacit knowledge (Polanyi, 1962) as knowledge beneath some level of describability or expressibility by extending unawareness structures (Heifetz, Meier, and Schipper, 2006, 2008) to multiple levels of awareness. In future work, we want to study to what extent implicit knowledge can satisfies less stringent properties than explicit knowledge (or vice versa) while still being coherent with each other. We like to explore whether such less stringent notions of implicit knowledge can be useful to rethink implicit biases (Brownstein, 2019) or implicit cognition in cognitive psychology (Augusto, 2010).


\end{document}